\documentclass[preprint]{revtex4}
\usepackage[utf8]{inputenc}
\usepackage{graphicx}
\usepackage{amsmath}
\usepackage{xcolor}
\usepackage{xspace}
\usepackage{amssymb}
\usepackage{tabularx}
\usepackage{ulem}
\usepackage{appendix}
\usepackage{pdflscape}
\usepackage{rotating}
\usepackage{appendix}

\def\meanXmax{\ifmmode {\langle X_\mathrm{max}\rangle}\else{$\langle X_\mathrm{max}\rangle$}\fi\xspace}%
\def\sigmaXmax{\ifmmode {\mathrm{RMS}(X_\mathrm{max})}\else{RMS$(X_\mathrm{max})$}\fi\xspace}%
\def\xmax{\ifmmode {X_\mathrm{max}}\else{$X_\mathrm{max}$}\fi\xspace}%
\newcommand{\sibyll}{{\scshape Sibyll2.3c}\xspace}
\newcommand{\epos}{{\scshape EPOS-LHC}\xspace}
\newcommand{\qgsjet}{{\scshape QGSJetII.04}\xspace}

\begin{document}

\newcolumntype{L}[1]{>{\raggedright\arraybackslash}p{#1}}
\newcolumntype{C}[1]{>{\centering\arraybackslash}p{#1}}
\newcolumntype{R}[1]{>{\raggedleft\arraybackslash}p{#1}}

\title{On the parametrization of the distributions of depth of shower maximum of ultra-high energy extensive air showers}
\author{Luan B. Arbeletche and  Vitor de Souza}
\affiliation{Instituto de F\'isica de S\~ao Carlos, Universidade de S\~ao Paulo, Av. Trabalhador S\~ao-carlense 400, S\~ao Carlos, Brasil.}
\date{\today}

\begin{abstract}
    The distribution of depth in which a cosmic ray air shower reaches its maximum number of particles (\xmax) is studied and parametrized. Three functions are studied for proton, carbon, silicon, and iron primary particles with energies ranging from $10^{17}\,$eV to $10^{20}\,$eV for three hadronic interaction models: \epos, \qgsjet, and \sibyll. The function which best describes the \xmax distribution of a mixed composition is also studied. A very large number of simulated showers and a detailed analysis procedure are used to guarantee negligible effects of undersampling and of fitting in the final results. For the first time, a comparison of several functions is presented under the same assumption with the intention of selecting within them the best functional form to describe the \xmax distribution. The Generalized Gumbel distribution is shown to be the best option among the options for a general description of all cases. The Log-normal distribution is also a good choice for some cases while the Exponentially Modified Gaussian distribution has shown to be the worst choice in almost all cases studied. All three functions are parametrized as a function of energy and primary mass.

\vspace{0.9cm}

\noindent
{\bf Note on this version:}
    an update to the parametrization discussed in Section~\ref{sec:parametrization} and whose parameters are tabulated in Appendix~\ref{app:param} has been added as Appendix~\ref{app:update}. This update envisages the inclusion of Helium primaries in the simulation library, which results in an improved description of the $X_\text{max}$ distributions for masses in the range $A = 1$ to $A = 12$. For details on the method and the functional form of the new parametrization, refer to Section~\ref{sec:parametrization}; these two aspects remain unchanged. The new values of the parametrization coefficients and their statistical uncertainties can be found in Appendix~\ref{app:update}. We thank Olena Tkachenko for calling our attention to the need of including helium primaries for a better description of all elements.
\end{abstract}

\maketitle

\section{Introduction}

The relative abundance of particle compositions in ultra-high energy cosmic rays (UHECR) is of key importance in the understanding of their acceleration mechanisms and interactions with extra-galactic radiation fields. The maximum particle energy of each source, the probability of escape from the acceleration region, the luminosity of the sources classes, the mean free path of the interaction on the way to Earth and the deviation angle in the magnetic fields are examples of fundamental astrophysics phenomena which depend on the particle type (mass and/or charge). A major improvement in our understanding of UHECR physics will not be possible without a precise determination of the abundance of each particle type. With this in mind, the two most important UHECR observatories (The Pierre Auger and the Telescope Array Observatories) are implementing upgrades to enhance their capabilities to determine the relative abundances of particle types arriving on Earth.

At these energies, the particles hitting Earth, called primary particles, are not directly observed. Their interaction with the atmosphere generates a cascade of particles which is measured by telescopes and ground detectors. The properties of the primary particle can be reconstructed from the detected signal of the shower. The most used and reliable parameter to determine the primary particle type is the depth at which the cascade reaches its maximum number of particles (\xmax). These extensive air showers are very complex branching processes whose stochastic behavior, although well understood in terms of particle interaction processes, cannot be solved analytically. Thus, fluctuations of important global quantities such as \xmax have no known functional form. In this sense, one always has to rely on Monte Carlo simulations to understand the intrinsic fluctuations of extensive air showers. Moreover, an approximation to the functional form of global variables can only be determined by the parametrization of simulated quantities.

Constant improvements in the fluorescence technique have allowed the Pierre Auger Observatory to measure \xmax with a systematic uncertainties of about $\pm 8\,$g/cm$^2$~\cite{auger-xmax} and the TA Collaboration quotes systematic uncertainties of $\pm 17.4\,$g/cm$^2$~\cite{ta-xmax}. The resolution in \meanXmax are quoted to be smaller than $25\,$g/cm$^2$ for the Pierre Auger Observatory measurements. The precision in measuring \xmax is such that new studies are based on the full distribution instead of only its moments~\cite{auger-xmax,ta-xmax,auger-vs-ta,mass-bellido-2018,bellido-xmax-2017}.

In this context, a good understanding of the \xmax distribution shape is mandatory since many steps in the analysis procedures depend on knowing \textit{a priori} its expected shape. The adoption of a particular parametrization may cause a wrong interpretation ox \xmax distributions when studying the primary fractions and their evolution with energy. Some functions have been proposed to describe the \xmax distribution~\cite{xmax1,xmax2} but no comparison between them is available. In this paper, three functions are used to describe the \xmax distribution and a detailed statistical comparison between them is presented. The purpose of this paper is to select the best description of the \xmax distribution and parametrize its dependencies with energy and mass.

This study is based on Monte Carlo simulations of air shower which are discussed in section~\ref{sec:sim}. In section~\ref{sec:functions}, the functions used to describe the \xmax distribution are presented and discussed. Section~\ref{sec:results} presents the results of the fits and section~\ref{sec:conclusion} presents the conclusions.

\section{Simulation of \xmax distributions}
\label{sec:sim}

Large samples of extensive air showers are simulated using the software \textsc{CONEX}~\cite{conex}. This software is an implementation of a one-dimensional hybrid model of the longitudinal development of particle cascades which has been extensively tested~\cite{xmax1}. Four atomic nuclei are considered: proton, carbon, silicon and iron (A = 1, 12, 28 and 56, respectively) with energies ranging from $10^{17}\,$eV to $10^{20}\,$eV in steps of 1 in $\log_{10}(E_0/\text{eV})$. The incident zenith angle of the primary cosmic rays is set to 75º. The longitudinal development is sampled in steps of $10\,$g/cm$^2$ until particles reach sea level, corresponding to a slant depth of $3860\,$g/cm$^2$. The energy cutoff for hadrons, muons, electrons, and photons is $1\,$GeV, $1\,$GeV, $1\,$MeV and $1\,$MeV, respectively. Given the known dependence on hadronic interaction models~\cite{rev-eas-had,syst,eas-diffrac}, three post-LHC hadronic interaction models are considered: \epos~\cite{epos}, \qgsjet~\cite{qgs2} and \sibyll~\cite{sib}. For each combination of primary mass, energy and hadronic interaction model, 10$^6$ showers are simulated.

\textsc{CONEX} provides the depth at which a shower reaches its maximum number of particles (\textsc{Xmx} variable in \textsc{CONEX} output) and the depth at which the energy deposit profile reaches its maximum (\textsc{XmxdEdX} in \textsc{CONEX} output). \textsc{Xmx} and \textsc{XmxdEdX} are calculated by fitting a quadratic function around the maximum of the longitudinal particle and energy deposit profile, respectively. Details of the fitting procedure can be found in the \textsc{CONEX} manual~\cite{conex}. These variables are compared and a maximum difference of $0.8 \pm 3.4\,$g/cm$^2$ between them is found in all simulated cases. Given that the difference between these variables is very small, much smaller than the uncertainties of the measurements, the depth at which the shower reaches the maximum number of particles (\textsc{Xmx}=\xmax) is used in the following calculations. The small difference and statistical uncertainty between \textsc{Xmx} and \textsc{XmxdEdX} also illustrate the quality of the fit done in \textsc{CONEX}.

Showers with two maxima in the longitudinal profile, the so-called double bump showers~\cite{anomalous}, for which the depth of shower maximum is not an unambiguously defined quantity, are removed from our analysis. The double bump showers are identified by searching the longitudinal profiles with more than two inflection points by computing the second derivative of profiles at each point in terms of finite differences. This approach effectively removes showers with two pronounced peaks. The fraction of removed profiles is below 0.4\% for all combinations of primary, energy and hadronic model.

\begin{figure}[]
  \centering
  \includegraphics[width=\textwidth]{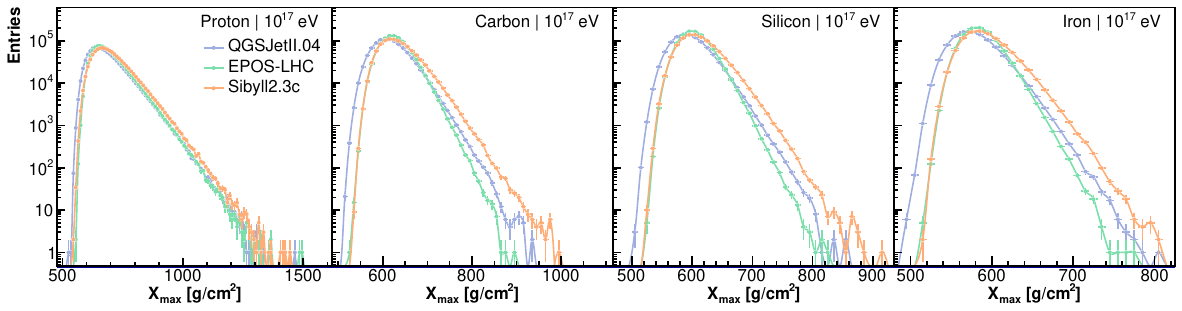}
  \includegraphics[width=\textwidth]{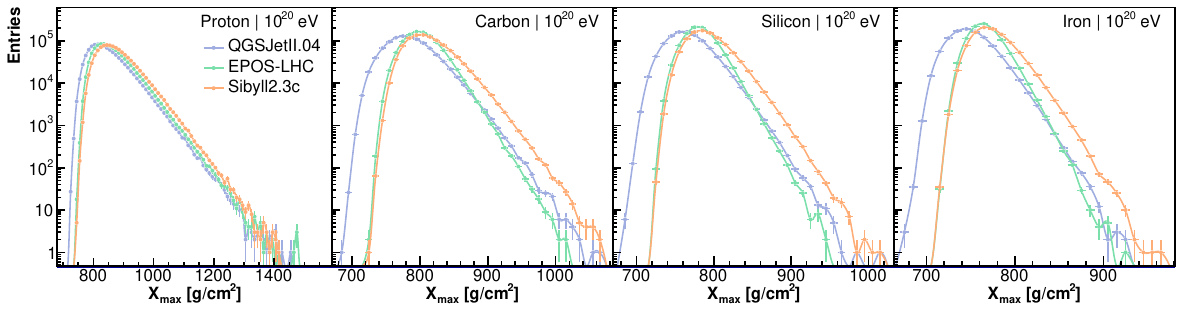}
  \caption{Examples of simulated \xmax distributions for four different primary masses (proton, carbon, silicon, iron) and three hadronic interaction models (\qgsjet,\epos and \sibyll) at the energies 10$^{17}\,$eV (upper panel) and 10$^{20}\,$eV (lower panel).}
  \label{fig:dist}
\end{figure}

Examples of simulated \xmax distributions for some primary masses with energies of $10^{17}\,$eV (upper panel) and $10^{20}\,$eV (lower panel) are shown in figure~\ref{fig:dist}. Primary types are indicated in the top-right corner of each plot. Each colored line corresponds to simulations done with a particular hadronic interaction model, indicated in the legend of the left plots. These distributions, as already known, have an accentuated positive skew that results from the exponential nature of particle interaction length distributions. Note, in figure~\ref{fig:dist}, the logarithm scale in the y-axis and the very small fluctuations of each point. In this illustration, \xmax was binned in intervals of $10\,$g/cm$^2$. As a result from the large simulated samples, fluctuations in the obtained distributions become larger only for very deep showers.

\section{Proposed functions to describe the \xmax distributions}
\label{sec:functions}

In this section, three functions are studied to parametrize the \xmax distributions: Exponentially Modified Gaussian, Generalized Gumbel, and Log-normal. They are going to be explained below and, whenever possible, an interpretation of their parameters is going to be given. Other functions have also been tested: Amoroso, Lévy $\alpha$-stable, Fréchet, Exponentiated Fréchet, Exponentiated Exponential, Landau and Gamma, but either they do not showed a good description of the \xmax distributions or they unreasonably increased the number of fitting parameters without providing a better description of data. The motivation for each function is also going to be briefly explored.

\subsection{Exponentially Modified Gaussian distribution}

The Exponentially Modified Gaussian (EMG) distribution was proposed in~\cite{xmax1} to describe \xmax distributions. It is motivated by the assumption that \xmax can be decomposed as $\xmax = X_\text{first} + \Delta \xmax$, where $X_\text{first}$ is the depth of the first interaction and $\Delta \xmax$ represents the shower development after the first interaction. While $X_\text{first}$ is known to have an exponential distribution with the mean free path $\lambda$, the distribution of $\Delta \xmax$ is unknown. The simplest approach is to assume that $\Delta \xmax$ is normally distributed with an average $\mu$ and variance $\sigma^2$, so that \xmax is distributed according to the convolution of an exponential and a Gaussian. The resulting function is:

\begin{equation}
  f(x) = \frac{1}{2\lambda}
  \exp \left( -\frac{x - \mu}{\lambda} + \frac{\sigma^2}{2\lambda^2} \right) \text{erfc} \left( \frac{\mu-x+\sigma^2/\lambda}{\sqrt{2}\sigma} \right) \, ,
  \label{eq:expgauss}
\end{equation}

\noindent where $\text{erfc}(x)$ is the complementary error function.

The EMG has three parameters that can be interpreted in terms of extensive air showers physics. $\lambda$, $\mu$ and $\sigma$ are related to the decay factor of the exponential, the depth of maximum of the distribution and the width of the distribution, respectively. $\sigma$ and $\lambda$ influence both the width and the mean of the \xmax distribution in different ways, mathematically $\langle X_{max} \rangle = \mu + \lambda$ and $\sigmaXmax = \sqrt{\sigma^2 + \lambda^2}$. The EMG function has already been employed in studies such as the determination of~\xmax moments from Pierre Auger Observatory~\cite{auger-xmax}, the comparison between Pierre Auger Observatory and Telescope Array \xmax data~\cite{auger-vs-ta} and the proposal of new methods to study the mass composition from real \xmax data~\cite{mass-bellido-2018}.

\subsection{Generalized Gumbel distribution}

The Gumbel distribution arises in the field of extreme value statistics to describe the frequency of extreme events (either minimum or maximum) in series of independent and identically distributed random variables~\cite{gumbel-book}. The Generalized Gumbel distribution (GMB) \cite{bertin-global-fluctuations} is written as

\begin{equation}
  f(x) = \frac{1}{\sigma} \frac{\lambda^\lambda}{\Gamma(\lambda)}
  \exp \left\{ -\lambda \left[ \frac{x-\mu}{\sigma} + \exp \left( -\frac{x-\mu}{\sigma} \right) \right] \right\} \, .
  \label{eq:gengumbel}
\end{equation}

\noindent Note that for $\lambda = 1$ one recovers the standard Gumbel distribution. Equation~\ref{eq:gengumbel} was proposed by reference~\cite{xmax2} to describe \xmax distributions.

The importance of the GMB distribution in extreme value statistics and its relation with the statistics of sums~\cite{stat-sum} can give some insight on its use to describe the \xmax distribution. Suppose a series of random variables $X_k$ is exponentially distributed according to

\begin{equation}
  g_k(x) = \frac{\lambda + k}{\sigma} \text{e}^{-(\lambda+k)x/\sigma} \, .
\end{equation}

\noindent It has been shown in reference~\cite{pink-noise} that the distribution of the sum $\sum_{k=0}^{\infty} X_k$ converges exactly to equation~\ref{eq:gengumbel} after a convenient shift and re-scaling of $X_k$. That is, the asymptotic sum of exponentially distributed random variables with increasing amplitudes converges to a GMB distribution. Based on it, it is possible to interpret \xmax as a sum of interaction depths of multiple generations of particles, similar to the model proposed in reference~\cite{matthews}, but with variable interaction lengths, and to write

\begin{equation}
  \xmax = \sum_{k=0}^{\eta - 1} X_k \, ,
  \label{eq:xmax-sum}
\end{equation}

\noindent where $\eta$ is the number of generations of particles. If $\eta \rightarrow \infty$ the distribution of the sum converges to equation~\ref{eq:gengumbel}. In this scenario, the mean free path of the first interaction is given by $\sigma/\lambda$. The scale parameter $\sigma$ describes how fast the average interaction lengths change between generations of particles. The location parameter ($\mu$) of equation~\ref{eq:gengumbel} is introduced to shift the mean of the distribution.

For finite $\eta$, the sum above follows a beta-exponential distribution~\cite{beta-exponential}:

\begin{equation}
  f(x) = \frac{1}{\sigma \text{B}(\eta,\lambda)} \text{e}^{-\lambda x/\sigma} \left( 1 - \text{e}^{-x/\sigma} \right)^{\eta-1} \, ,
  \label{eq:betaexp}
\end{equation}

\noindent where $\text{B}(x,y)$ is the beta function, defined for $x , y\geq0$. If a location parameter ($\mu$) is added to the beta-exponential distribution, it could as well be considered a candidate to describe \xmax distributions. The beta-exponential distribution was also tested following the method explained below, however, it did not show any improvement in the description of \xmax distribution in comparison to the GMB. Since the beta-exponential function has one parameter more than the GMB, it was decided to keep only the GMB for further studies which in total has also three parameters.

\subsection{Log-normal distribution}

The log-normal distribution is characteristic of stochastic processes where the variable of interest can be written as a product of independent and identically distributed random variables so that its logarithm is normally distributed according to the central limit theorem. The log-normal distribution (LOG) proves to be difficult to interpret in terms of extensive air showers. However, as it will be shown later, it provides a good description of \xmax distributions. The probability density function is given by

\begin{equation}
  f(x) =
  \begin{cases}
    0 \, , &\text{ if } x \leq m \\
    \frac{1}{\sqrt{2\pi}\sigma}\frac{1}{x-m}\exp\left\{  - \frac{\left[ \ln(x-m) - \mu \right]^2} {2 \sigma^2} \right\} \, , &\text{ if } x > m \, .
  \end{cases}
  \label{eq:lognormal}
\end{equation}

It has three parameters $m$, $\mu$ and $\sigma$ related to the position of the peak, the width of the distribution and the length of the tail, respectively.

\section{Fitting the \xmax distributions}
\label{sec:results}

The \xmax distributions of each combination of primary mass, energy and hadronic interaction model are fitted using the three functions presented in the previous sections. The best description of the \xmax distributions is achieved by searching for the three parameters in each function which resulted in the maximum likelihood. The \xmax distributions are not binned (unbinned fit). The \textsc{Minuit}~\cite{minuit} library available within the \textsc{ROOT} analysis framework~\cite{root} is used in the fitting procedure.

\begin{figure}[h!]
  \centering
  \includegraphics[width=0.44\textwidth]{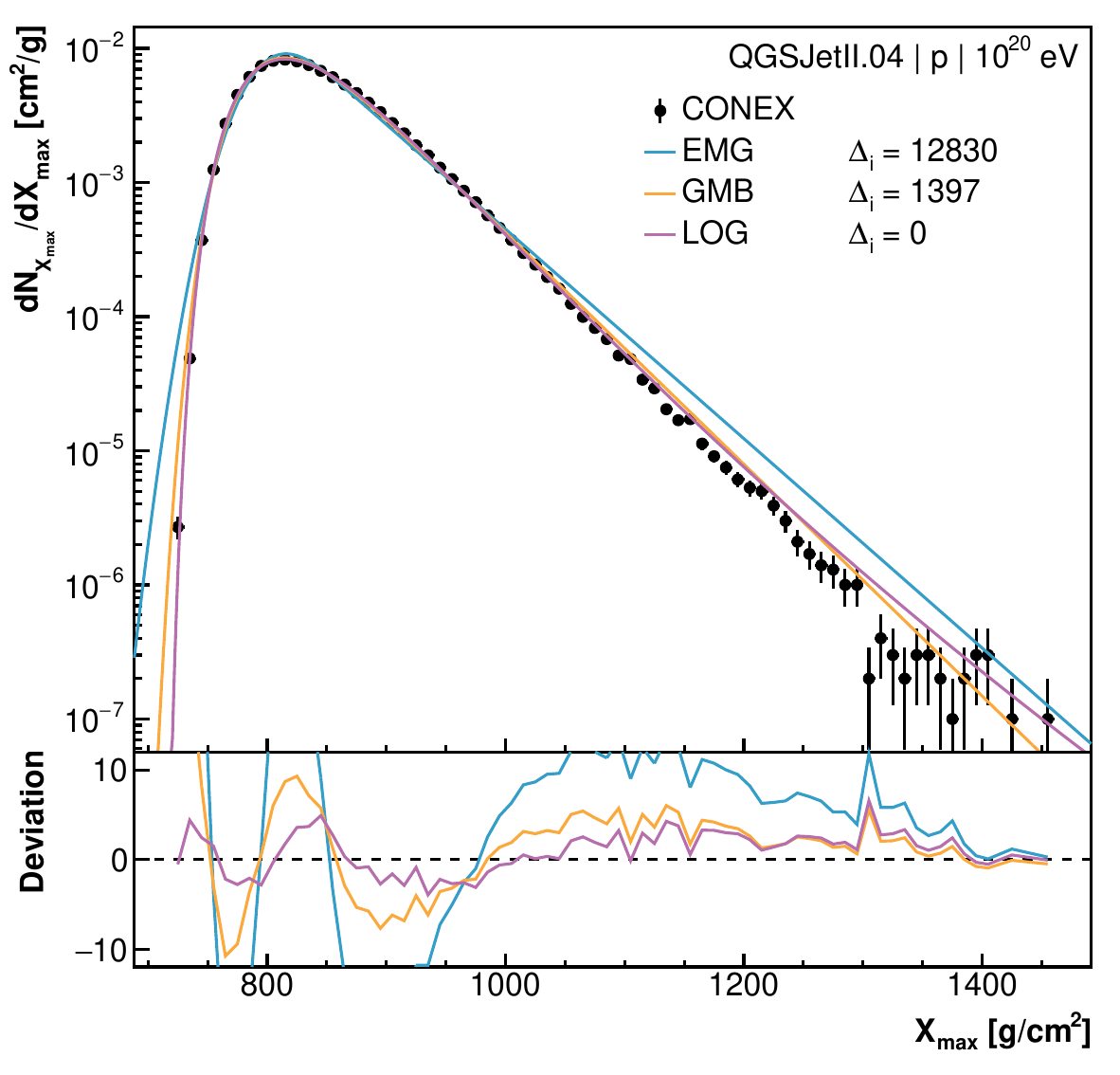}
  \includegraphics[width=0.44\textwidth]{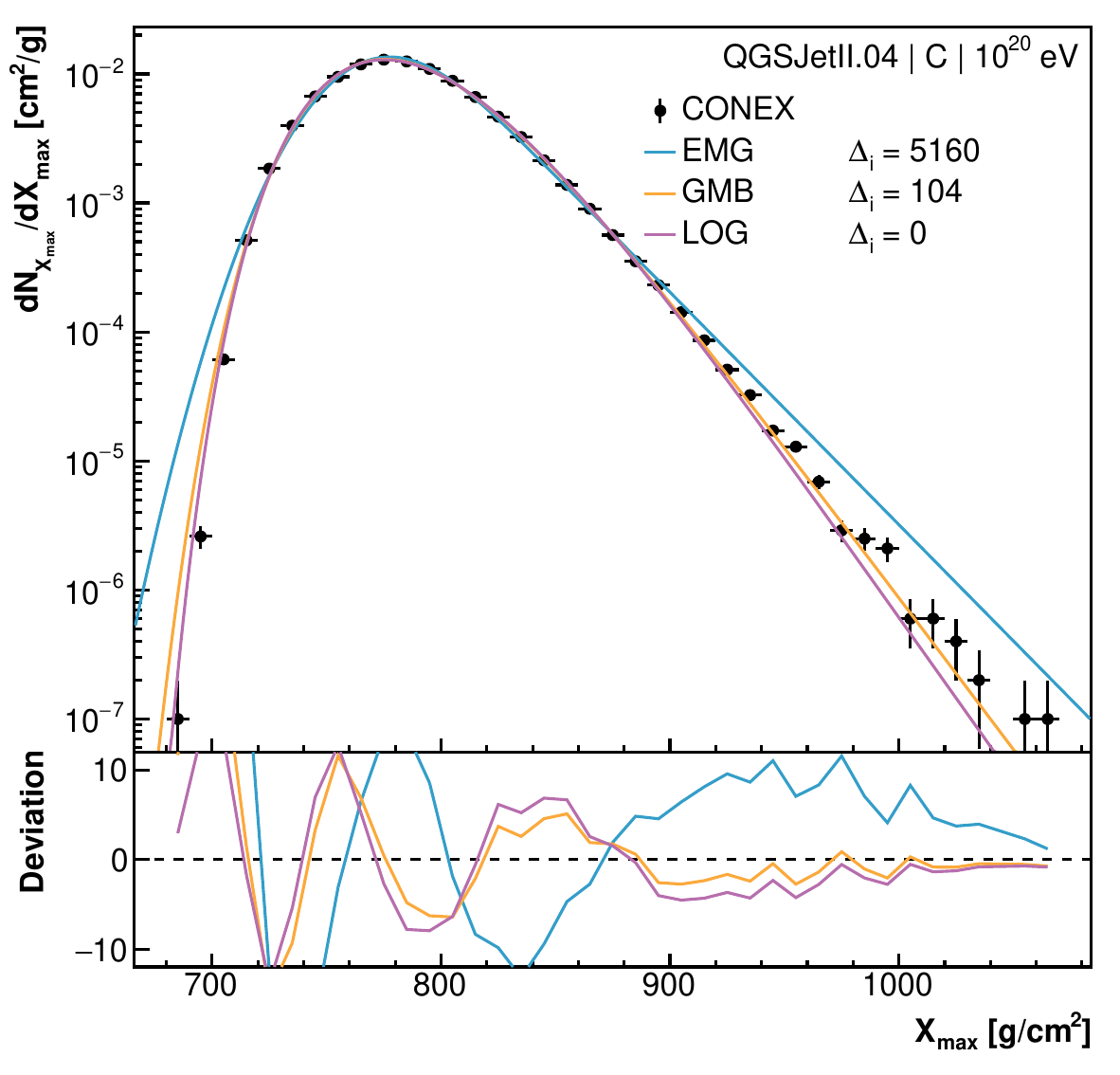} \\
  \includegraphics[width=0.44\textwidth]{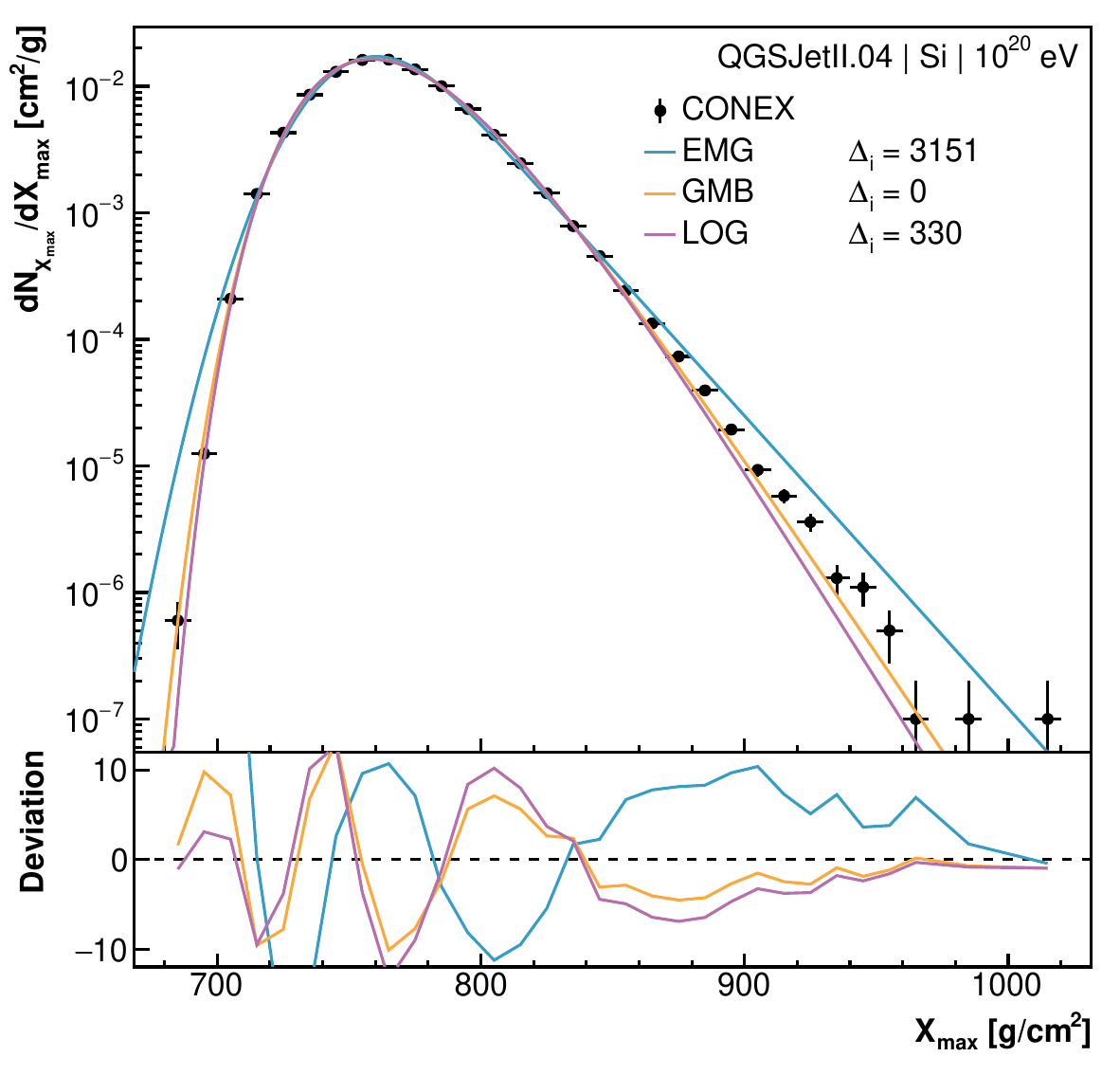} \includegraphics[width=0.44\textwidth]{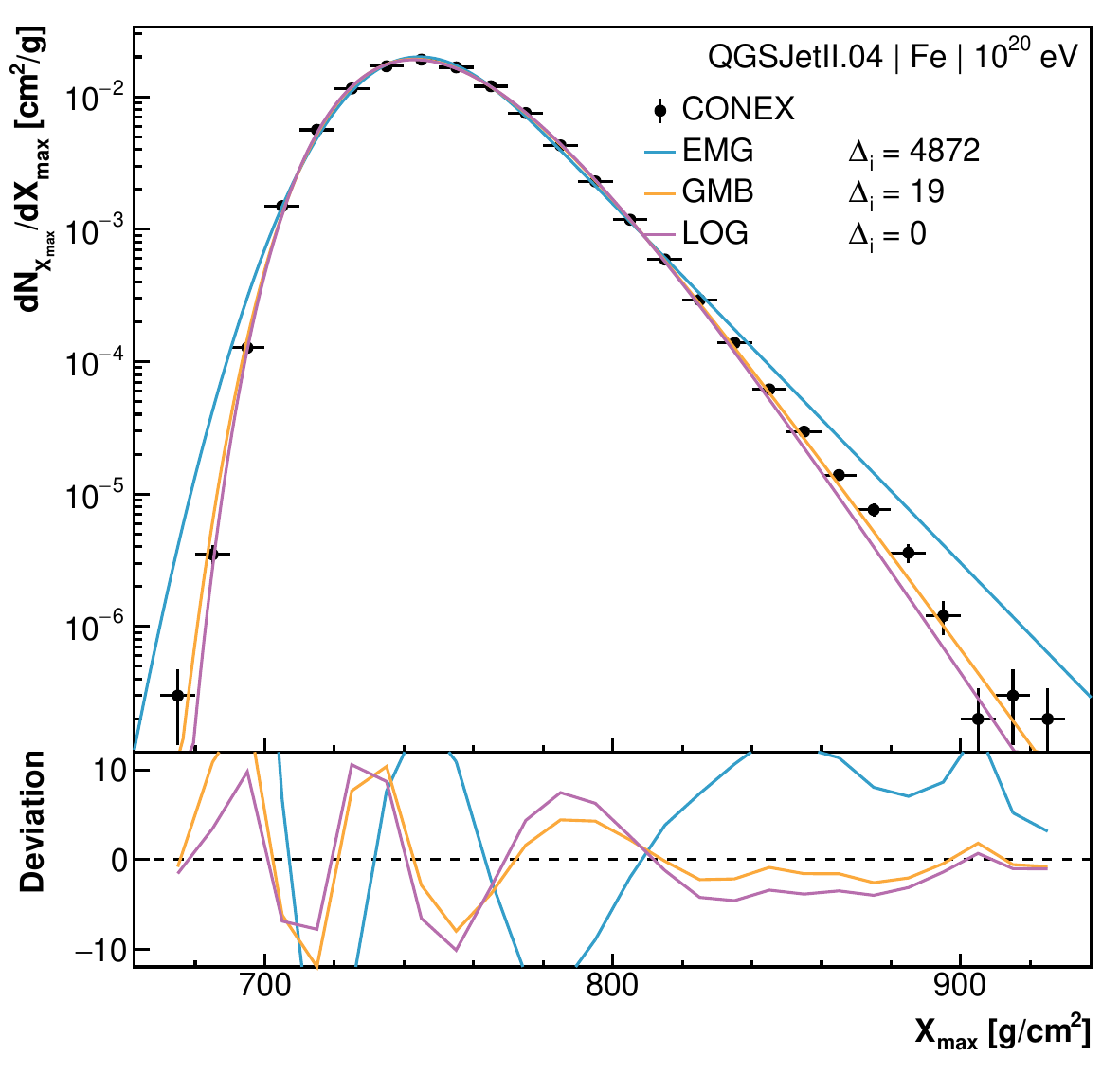}
  \caption{Examples of fits of \xmax distributions. The primary particle is indicated at the top-right corner of each plot. Fit functions are shown as colored solid lines, while the simulated \xmax distribution is shown as circular dots. The bottom panels show the deviation of each fitted function to the simulated point, defined as the difference between the function and the point divided by the statistical uncertainty of the point. Only results for \qgsjet are shown in this example.}
  \label{fig:fits}
\end{figure}

Examples of fitting results are presented in figure~\ref{fig:fits} for simulations obtained with \qgsjet at an energy of 10$^{20}\,$eV. Only for illustration purposes, the distributions are binned in intervals of $10\,$g/cm$^2$. Note the logarithmic scale in the y-axis. The primary particle is indicated at the top-right corner of each plot. Fit functions are shown as colored solid lines, while the simulated \xmax distribution is shown as circular dots. The bottom panels show the deviation (pull values) of each fitted function to the simulated point, defined as the difference between the function and the point divided by the statistical uncertainty of the point.

Figure~\ref{fig:fits} shows that the EMG distribution is not able to describe the simulated distributions for small and large \xmax values. No clear preference between the GMB and the LOG distributions is seen.

Values of the Akaike Information Criterion (AIC) for each case are shown in table~\ref{tab:comp}. Since the absolute value of the AIC has no meaning in this unbinned fit, the values shown are relative ($\Delta_i$) to the smallest AIC in each case. Reader is referred to appendix~\ref{app:aic} for the definition of AIC.

The first notable fact in table~\ref{tab:comp} is that the EMG distribution has the worst AIC value for every primary, energy and hadronic interaction model except one: silicon - 10$^{20}\,$eV - \epos for which the AIC value is slightly better than the GMB fit. This makes the EMG distribution the worst selection among the three functions described here to describe \xmax distributions of single primary particles.

The GMB and LOG distributions represent similar good description of the \xmax distributions. The LOG distribution performs better for low mass primaries (proton and carbon) and the GMB distribution performs better for heavier primaries (silicon and iron). But the differences between the quality of the fit of GMB and LOG are only marginal.

\begin{table}[h!]
  \centering
  \caption{Relative AIC values ($\Delta_i$) of the fit of the unbinned \xmax distributions for the three hadronic interaction models and primary particle energy ranging from $10^{17}\,$eV to $10^{20}\,$eV. Note that a value of zero for a model means that this is the best model for the respective energy, mass and hadronic model combination.}
  \scriptsize
  \begin{tabular}{c | C{0.78cm} C{0.78cm} C{0.78cm} C{0.78cm} | C{0.78cm} C{0.78cm} C{0.78cm} C{0.78cm} | C{0.78cm} C{0.78cm} C{0.78cm} C{0.78cm} | C{0.78cm} C{0.78cm} C{0.78cm} C{0.78cm}}
    \hline
    \hline
    \multicolumn{17}{c}{\qgsjet} \\
    \hline
    Primary & \multicolumn{4}{c|}{Proton} & \multicolumn{4}{c|}{Carbon} & \multicolumn{4}{c|}{Silicon} & \multicolumn{4}{c}{Iron}\\
    \hline
    $\log(E_0/\text{eV})$  &      17 &      18 &      19 &      20 &      17 &      18 &      19 &      20 &      17 &      18 &      19 &      20 &      17 &      18 &      19 &      20 \\
    \hline
    EMG &   10113 &   11209 &   12226 &   12830 &    6636 &    6099 &    5181 &    5160 &    4213 &    3743 &    3447 &    3151 &    4920 &    5251 &    4875 &    4872 \\
    GMB &     675 &    1044 &    1285 &    1397 &     131 &     105 &      32 &     104 &       0 &       0 &       0 &       0 &       0 &       0 &       0 &      19 \\
    LOG &       0 &       0 &       0 &       0 &       0 &       0 &       0 &       0 &     402 &     381 &     384 &     330 &     202 &      79 &      81 &       0 \\
    \hline
    \multicolumn{17}{c}{\epos} \\
    \hline
    EMG &    8932 &   10507 &   13115 &   14264 &    4325 &    3884 &    3728 &    3027 &    2156 &    1236 &    1315 &       0 &    1742 &    1066 &    1571 &    1563 \\
    GMB &      28 &     573 &    1425 &    1865 &       0 &       0 &       0 &       0 &       0 &       0 &       0 &     475 &       0 &       0 &       0 &       0 \\
    LOG &       0 &       0 &       0 &       0 &     232 &     293 &     262 &     272 &     526 &     629 &     643 &    1222 &     681 &     754 &     781 &     802 \\
    \hline
    \multicolumn{17}{c}{\sibyll} \\
    \hline
    EMG &    9319 &   10117 &   11619 &   12648 &   11851 &   11493 &   11277 &   10987 &    6492 &    6637 &    6559 &    6269 &    6542 &    6282 &    5655 &    4954 \\
    GMB &     420 &     666 &    1103 &    1362 &     914 &     805 &     760 &     713 &       0 &       0 &       0 &       0 &       0 &       0 &       0 &       0 \\
    LOG &       0 &       0 &       0 &       0 &       0 &       0 &       0 &       0 &     247 &     182 &     123 &     139 &     326 &     379 &     495 &     538 \\
    \hline
    \hline
  \end{tabular}
  \label{tab:comp}
\end{table}

\subsection{Mixed composition}

\begin{figure}
  \centering
  \caption{Example of \xmax distributions at an energy of $10^{18}\,$eV following the fractions shown in table~\ref{tab:fractions}. Filled histograms with color lines represent the distribution of each primary particle. Black dots shows the sum of all particles. Color lines shows the result of the fit of proposed functions to the distribution of all particles (black dots). Left panel for \epos and right panel for \sibyll.}
  \includegraphics[width=0.48\textwidth]{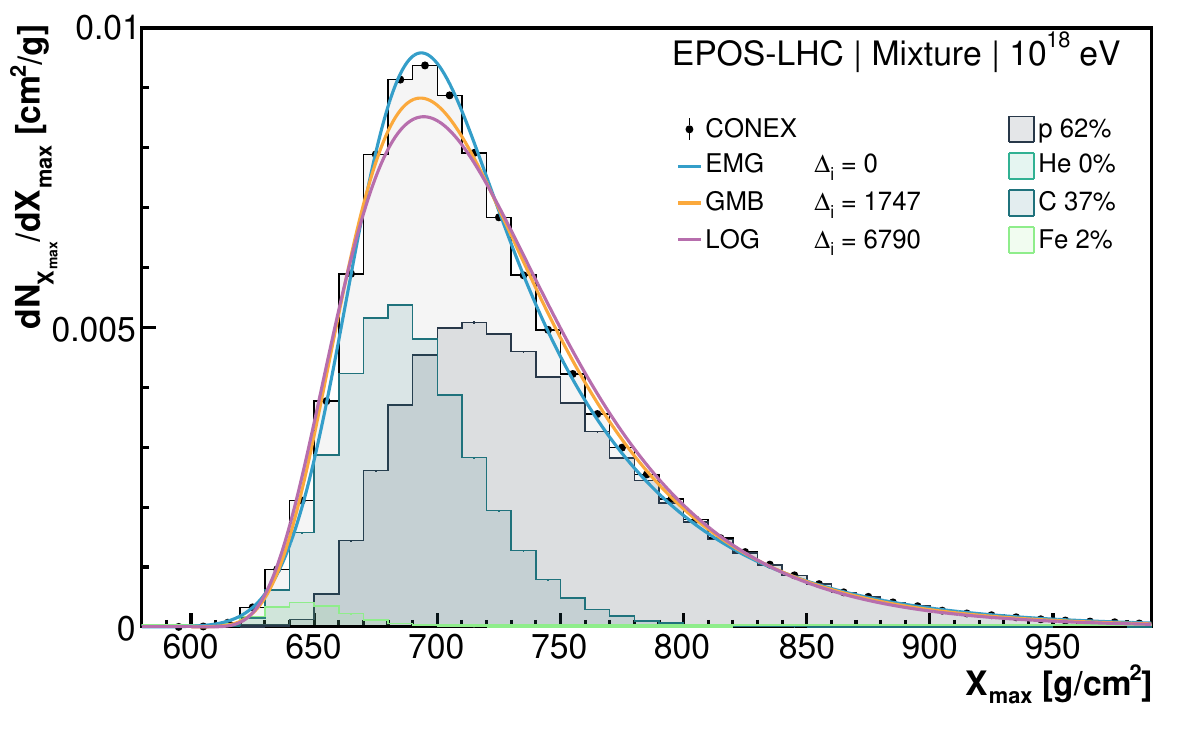}
  \includegraphics[width=0.48\textwidth]{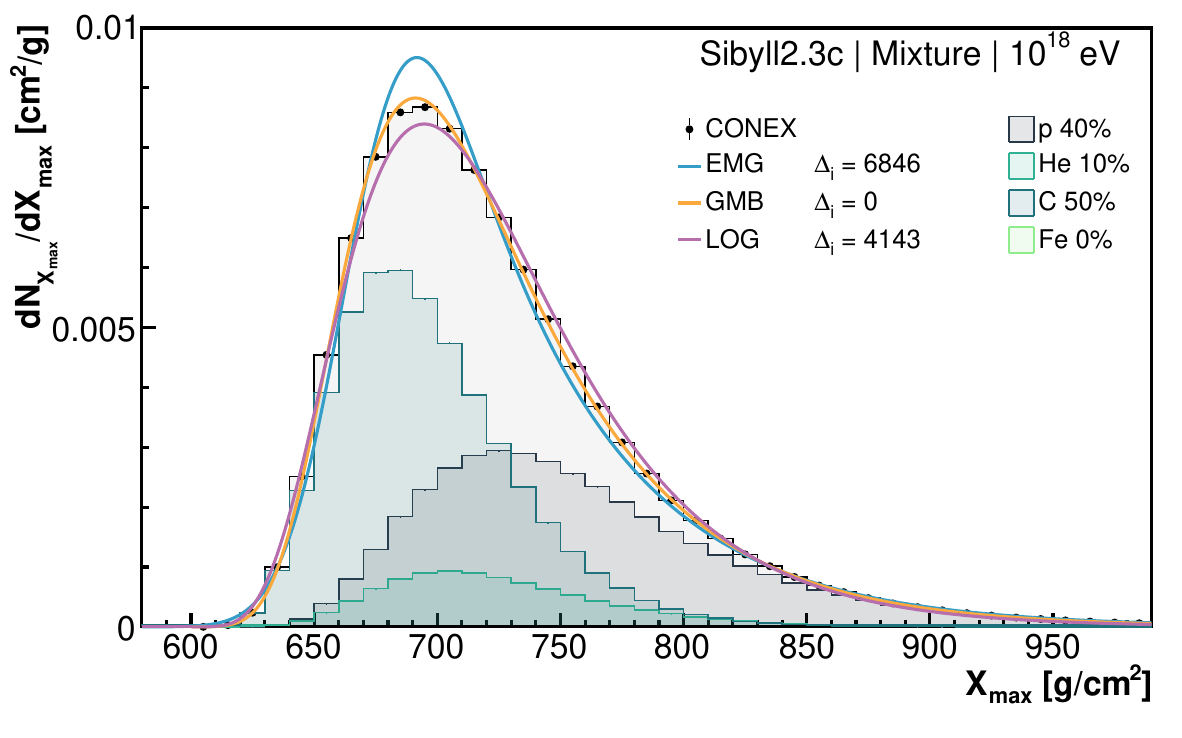}
  \label{fig:mixed}
\end{figure}

The \xmax distributions of events with energy between $10^{18}\,$eV and $10^{19}\,$eV measured by the Pierre Auger Observatory can be better described by a combination of primary particles rather than a pure element~\cite{bellido-xmax-2017,auger-comp-2014}. Therefore in the analysis of these distributions it is important that the used function is able to describe also mixes compositions instead of only pure samples. In this section, the proposed functions are going to be tested against a mix of primary particles. The simulated \xmax distributions are mixed following the fraction which best describes the data of the Pierre Auger Observatory as shown in reference~\cite{bellido-xmax-2017} and table~\ref{tab:fractions}. Figure~\ref{fig:mixed} shows two examples of mixtures at $10^{18}\,$eV for \epos and \sibyll models. Distributions are binned in intervals of $10\,$g/cm$^2$ for illustration purposes. The resulting mixture is fitted by the three proposed functions.

The $\Delta_i$ values for the fits are shown in table~\ref{tab:comp_mix}. The GMB shows an overall better description of the distributions, losing only marginally to the EMG for \epos at $10^{18}\,$eV and LOG for \sibyll at $10^{19}\,$eV.

\begin{table}[h!]
  \centering
  \footnotesize
  \begin{tabular}{c | C{1.5cm} | C{1.5cm} | C{1.5cm} | C{1.5cm} | C{1.5cm} | C{1.5cm}}
    \hline
    \hline
    Model& \multicolumn{2}{c |}{\epos} & \multicolumn{2}{c |}{\qgsjet} & \multicolumn{2}{c}{\sibyll} \\
    \hline
    $\log(E_0/\text{eV})$ & 18 & 19 & 18 & 19 & 18 & 19 \\
    \hline
    p  & 61.5\% &  9.5\% & 63.2\% & 35.6\% & 40.4\% &  2.8\% \\
    He &  0.0\% & 62.0\% & 36.8\% & 64.4\% &  9.7\% & 38.7\% \\
    C  & 36.7\% & 28.5\% &  0.0\% &  0.0\% & 49.9\% & 58.5\% \\
    Fe &  1.8\% &  0.0\% &  0.0\% &  0.0\% &  0.0\% &  0.0\% \\
    \hline
    \hline
  \end{tabular}
  \caption{Primary fractions which best describes the \xmax distributions measured by the Pierre Auger Observatory~\cite{bellido-xmax-2017} at the energies used in this paper.}
  \label{tab:fractions}
\end{table}

\begin{table}[h!]
  \centering
  \caption{Relative AIC values ($\Delta_i$) of the fit of the unbinned \xmax distributions for the three hadronic interaction models and mix of primary particle.
  Note that a value of zero for a model means that this is the best model for the respective energy, mass and hadronic model combination.}
  \footnotesize
  \begin{tabular}{c | C{1.5cm} | C{1.5cm} | C{1.5cm} | C{1.5cm} | C{1.5cm} | C{1.5cm}}
    \hline
    \hline
    Model& \multicolumn{2}{c |}{\epos} & \multicolumn{2}{c |}{\qgsjet} & \multicolumn{2}{c}{\sibyll} \\
    \hline
    $\log(E_0/\text{eV})$ & 18 & 19 & 18 & 19 & 18 &19 \\
    \hline
    EMG  & 0    & 5557 & 6093 & 5415 & 6846 & 9378 \\
    GMB  & 1747 & 0    &  0   &  0   & 0    &  200 \\
    LOG  & 6790 & 991  & 4813 & 1457 & 4143 & 0 \\
    \hline
    \hline
  \end{tabular}
  \label{tab:comp_mix}
\end{table}

\section{Parametrization of \xmax distributions as a function of energy and mass}
\label{sec:parametrization}

The three proposed functions are used to fit the simulated \xmax distributions for proton, carbon, silicon and iron with energies ranging from $10^{17}\,$eV to $10^{20}\,$eV in steps of 1 in $\log_{10}(E_0/\text{eV})$. Each function has three parameters as shown in section~\ref{sec:functions}. These parameters are modeled as a function of primary energy and mass. The proposed functional form is:

\begin{equation}
  \label{eq:p1}
  \theta(E_0,A) = a(A) + b(A) \log_{10} E_0 + c(A) (\log_{10} E_0)^2 \, ,
\end{equation}

\noindent where

\begin{align}
  \label{eq:p2}
  a(A) &= a_0 + a_1 \log_{10} A + a_2 (\log_{10} A)^2 \, , \nonumber \\ 
  b(A) &= b_0 + b_1 \log_{10} A + b_2 (\log_{10} A)^2 \, ,\\
  c(A) &= c_0 + c_1 \log_{10} A + c_2 (\log_{10} A)^2 \, . \nonumber
\end{align}

\noindent Values obtained for the parameters $a_i$, $b_i$ and $c_i$ and their corresponding statistical uncertainties are found in appendix~\ref{app:param}. Note that a value of zero in  table~\ref{tab:params} means the inclusion of that parameters leads to a worse fit of the simulated distribution.

\begin{figure}
  \centering
  \includegraphics[width=0.32\textwidth]{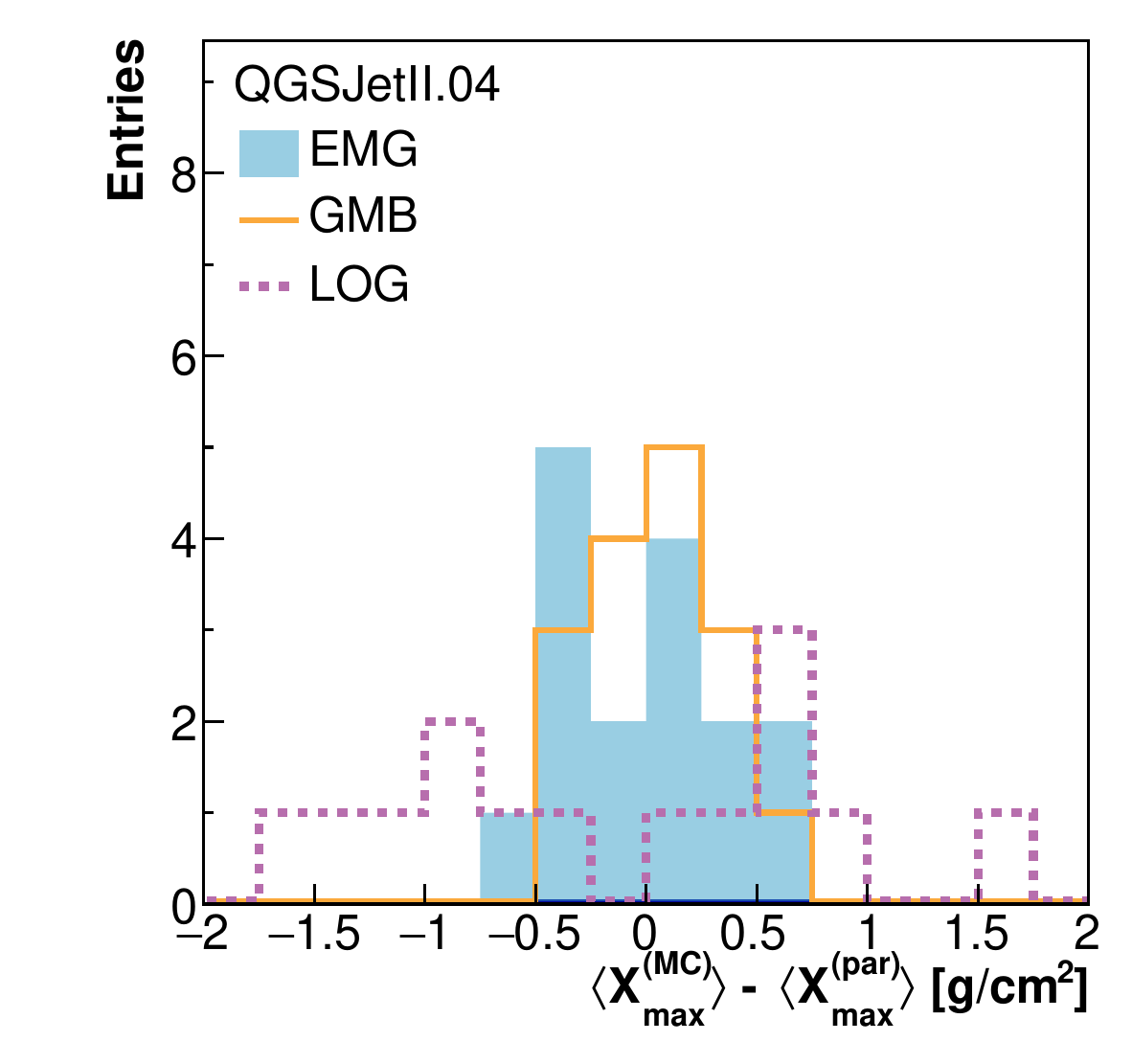}
  \includegraphics[width=0.32\textwidth]{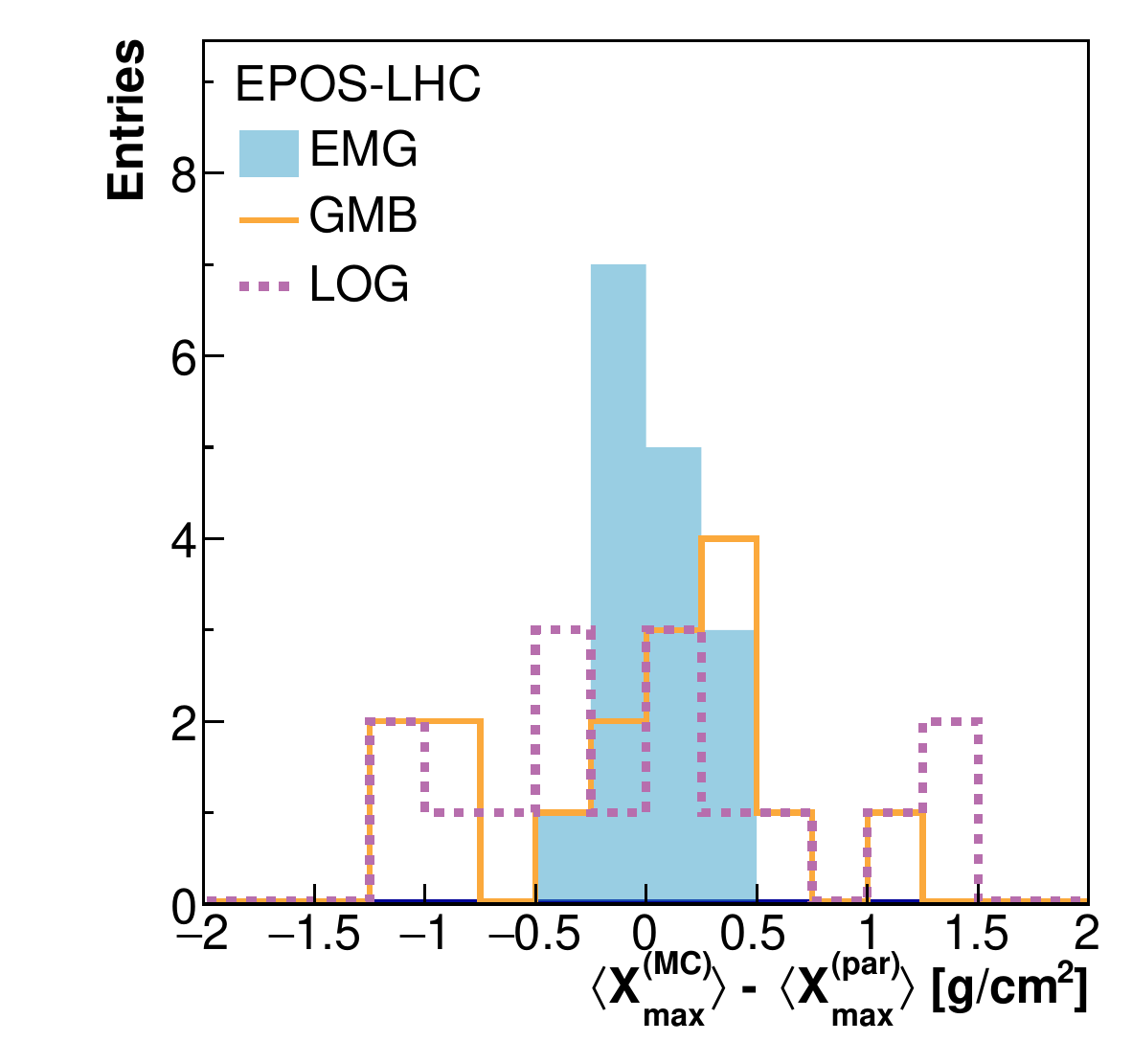}
  \includegraphics[width=0.32\textwidth]{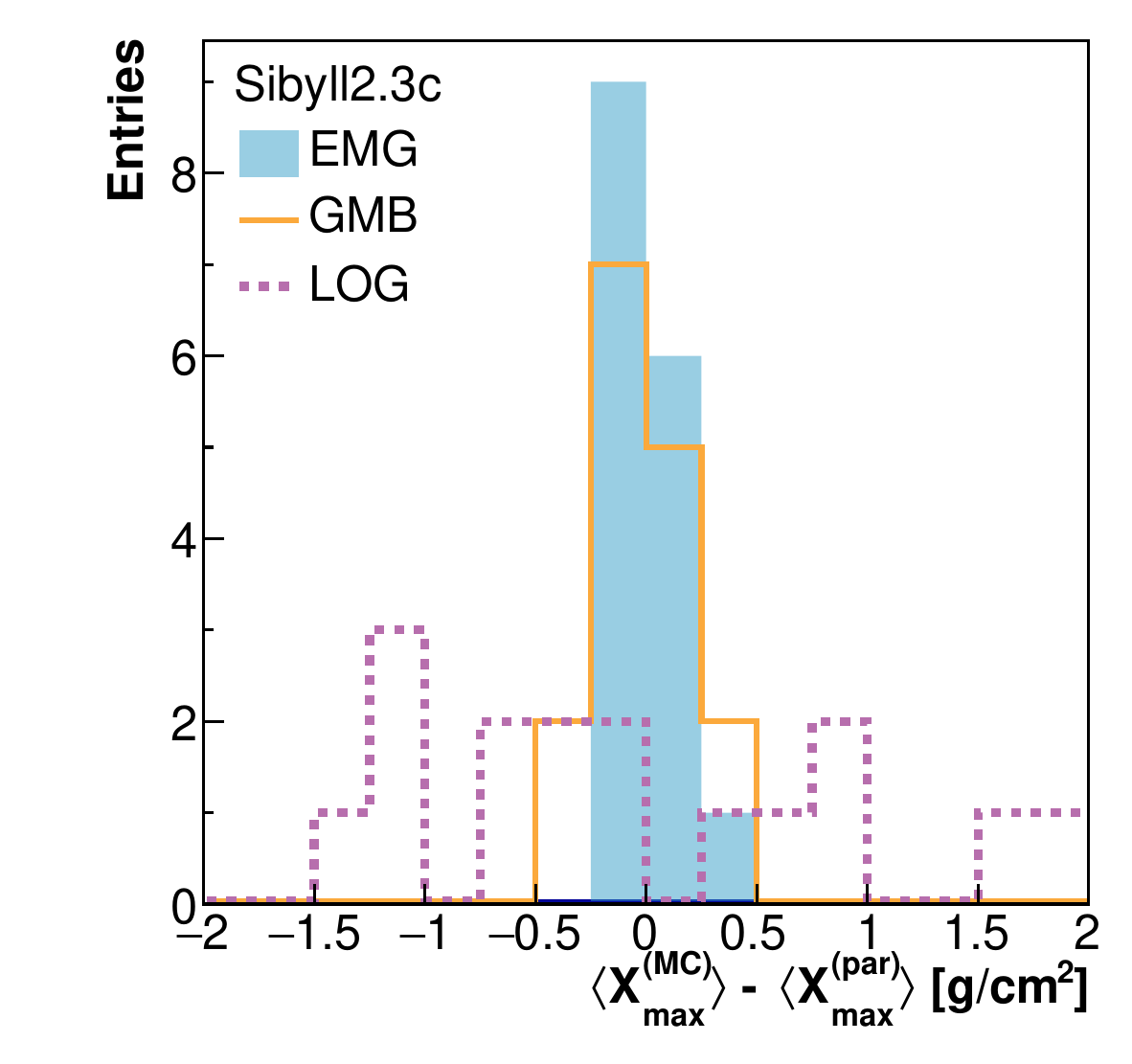} \\
  \includegraphics[width=0.32\textwidth]{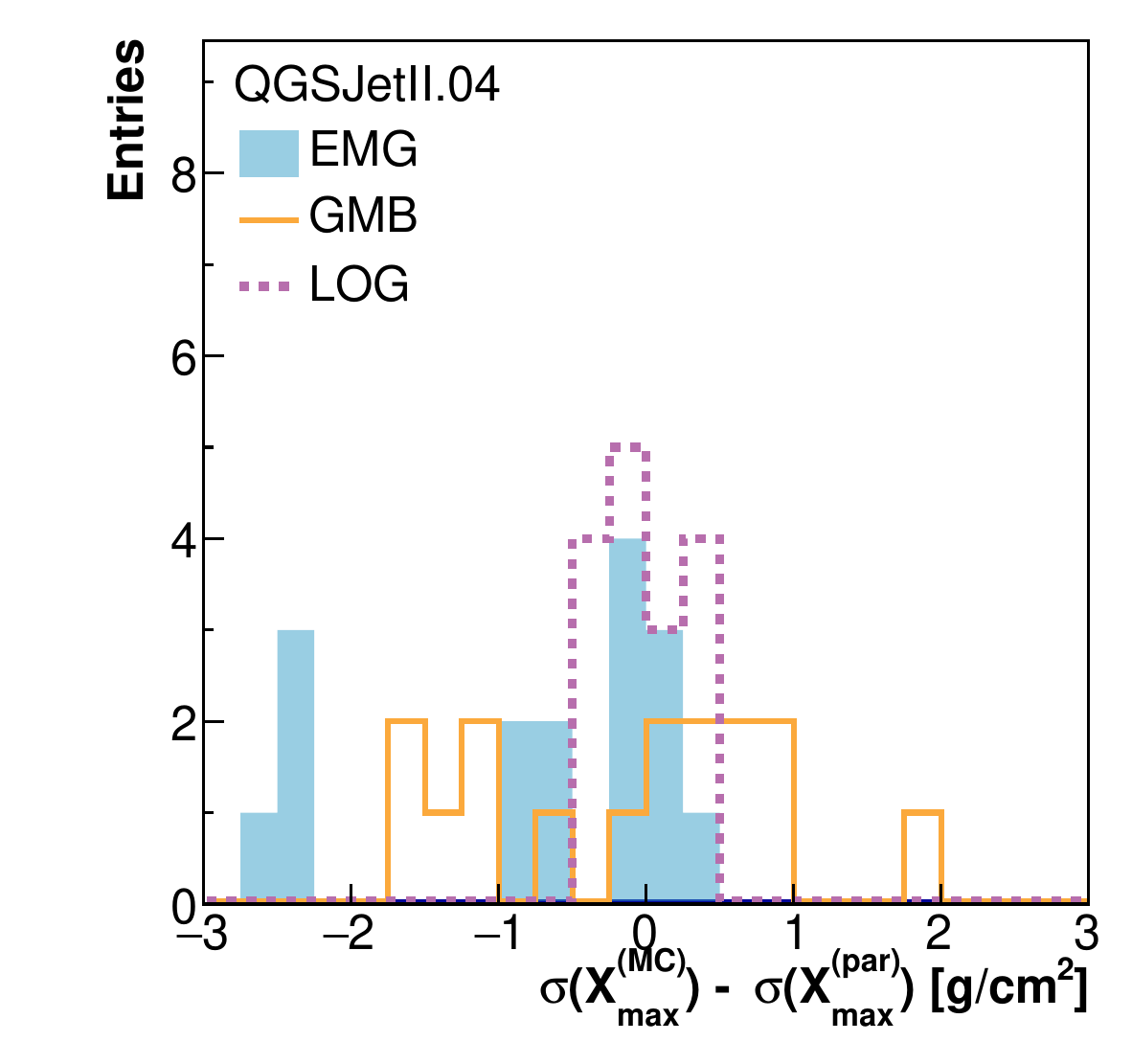}
  \includegraphics[width=0.32\textwidth]{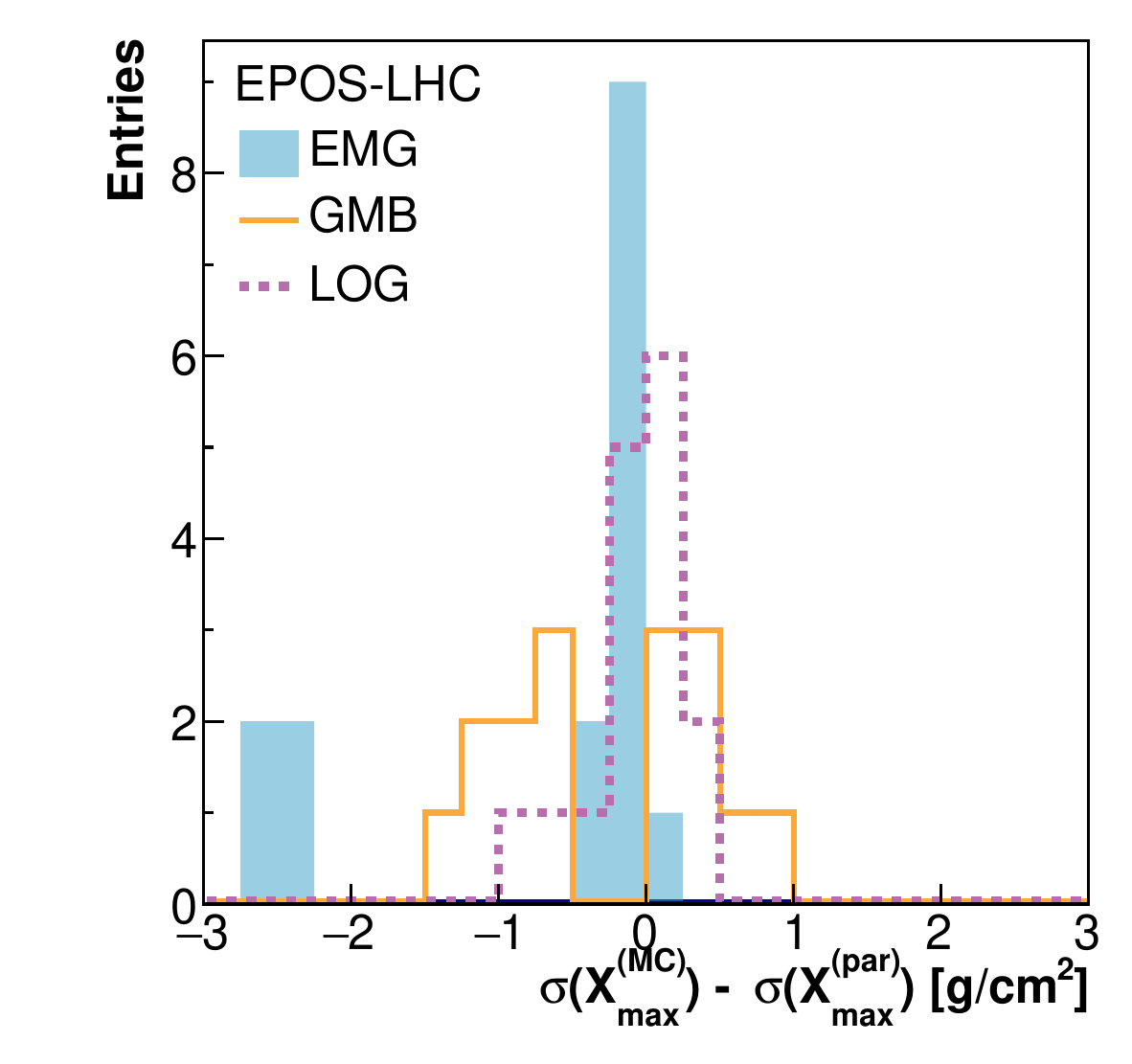}
  \includegraphics[width=0.32\textwidth]{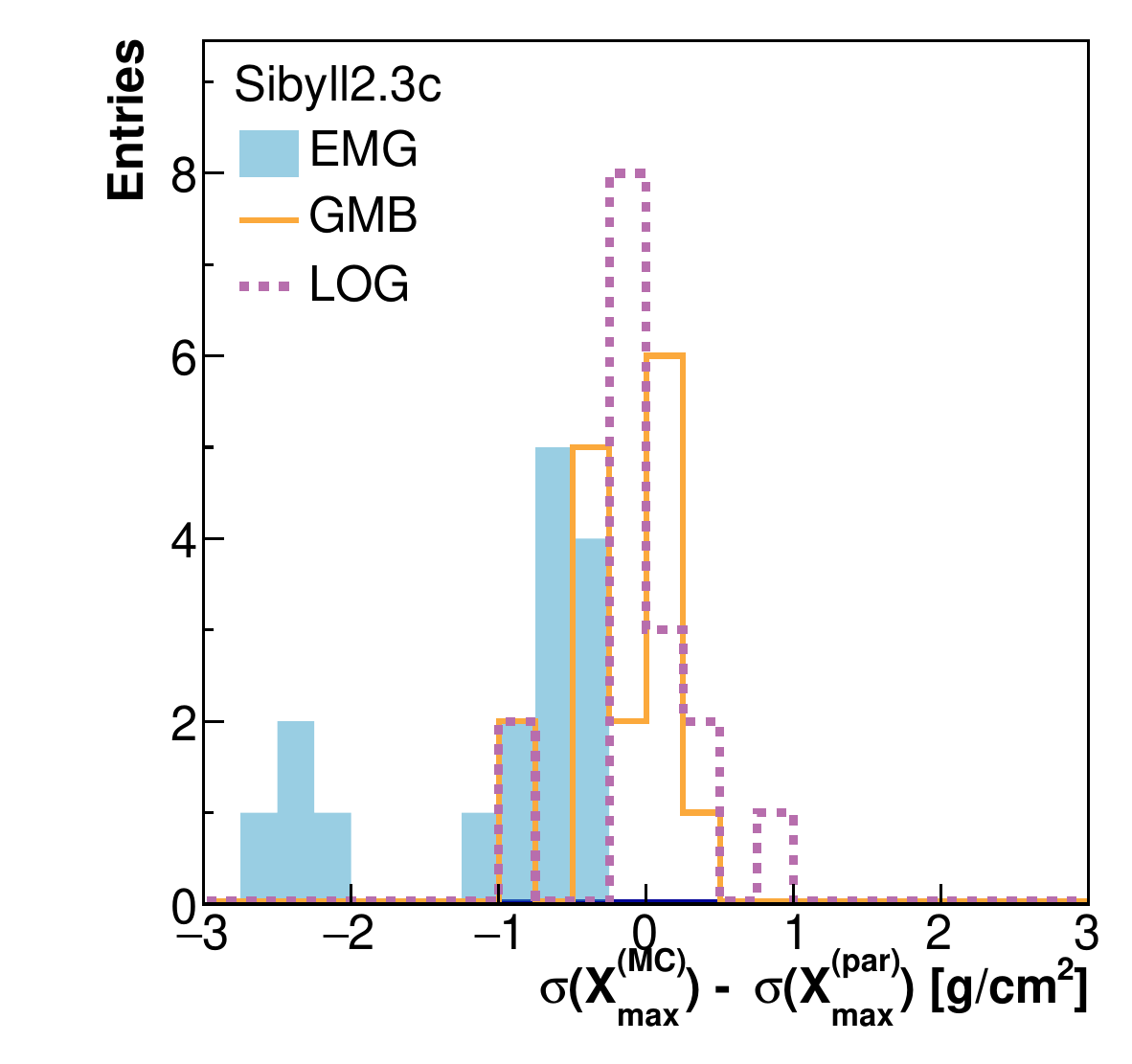}
  \caption{Error on the first moment (upper plots) and second moment (lower plots) between the parametrized distributions (par) and the simulated (MC) \xmax distributions.}
  \label{fig:compare}
\end{figure}

The error caused by the use of equations~\ref{eq:p1} and~\ref{eq:p2} to calculate the parameters as a function of mass and energy was determined by evaluating the differences between the first and second moments of the parametrized distributions and the simulated distributions for each mass and energy. Results are shown as histograms in figure~\ref{fig:compare}. The upper plots show the deviations on the first moment of the \xmax distributions for each hadronic interaction model indicated in the top left corner of each box. The lower plots show the differences for the second moment of the \xmax distributions. The largest difference between the proposed parametrization and the simulations is $2\,$g/cm$^2$ for the first moment and $3\,$g/cm$^2$ for the second moment.

\begin{figure}
  \centering
  \includegraphics[width=0.32\textwidth]{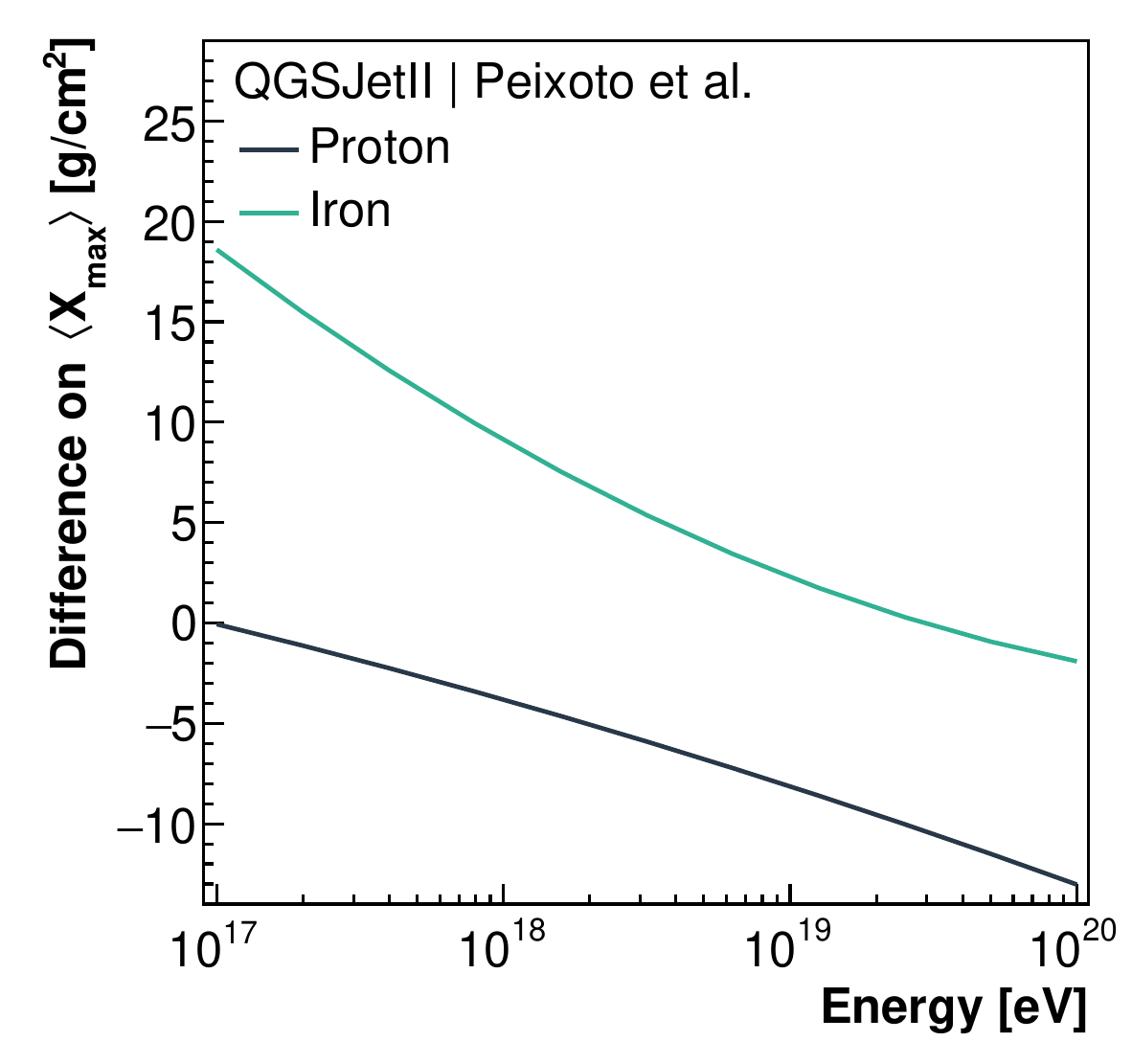}
  \includegraphics[width=0.32\textwidth]{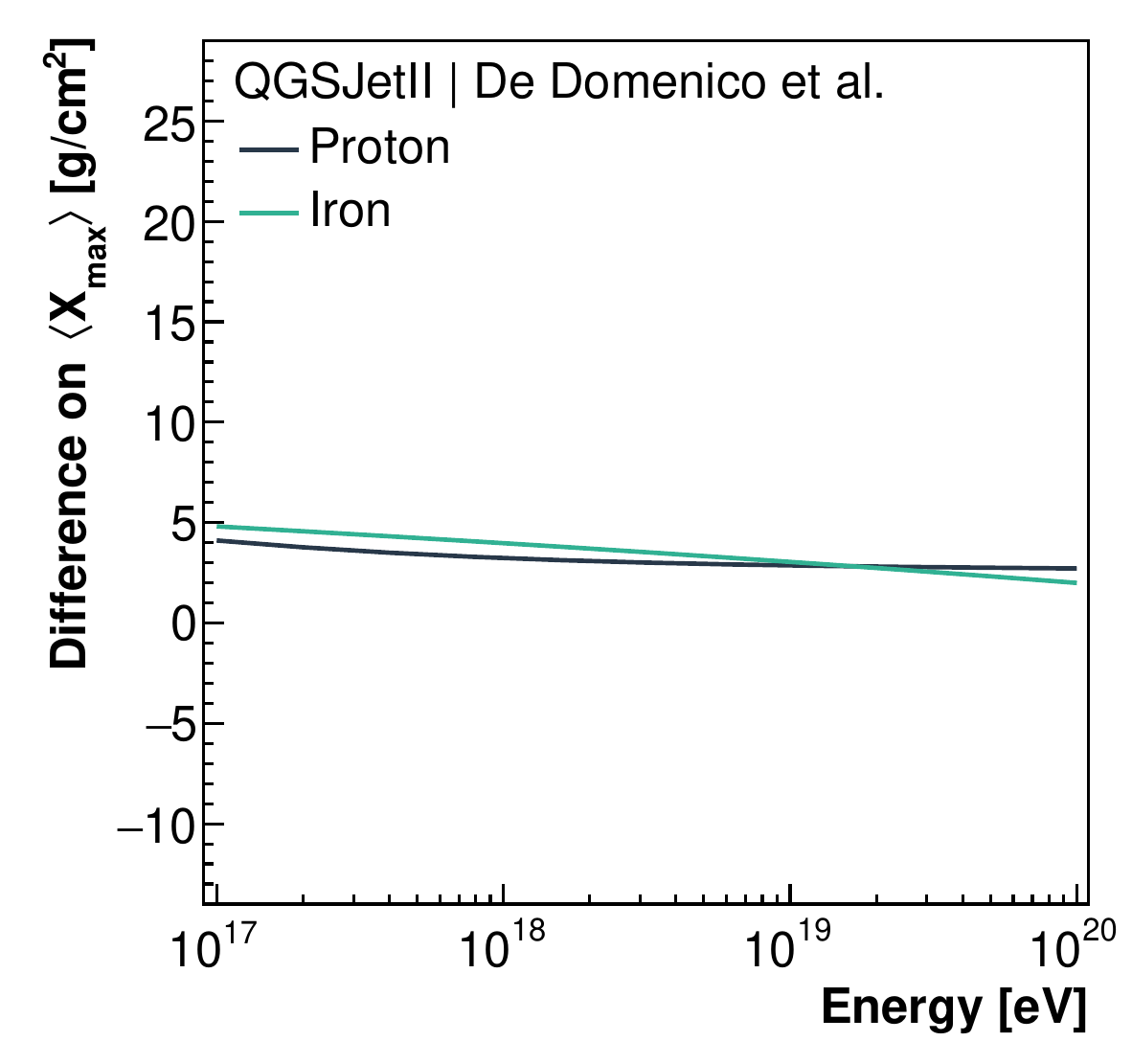}
  \includegraphics[width=0.32\textwidth]{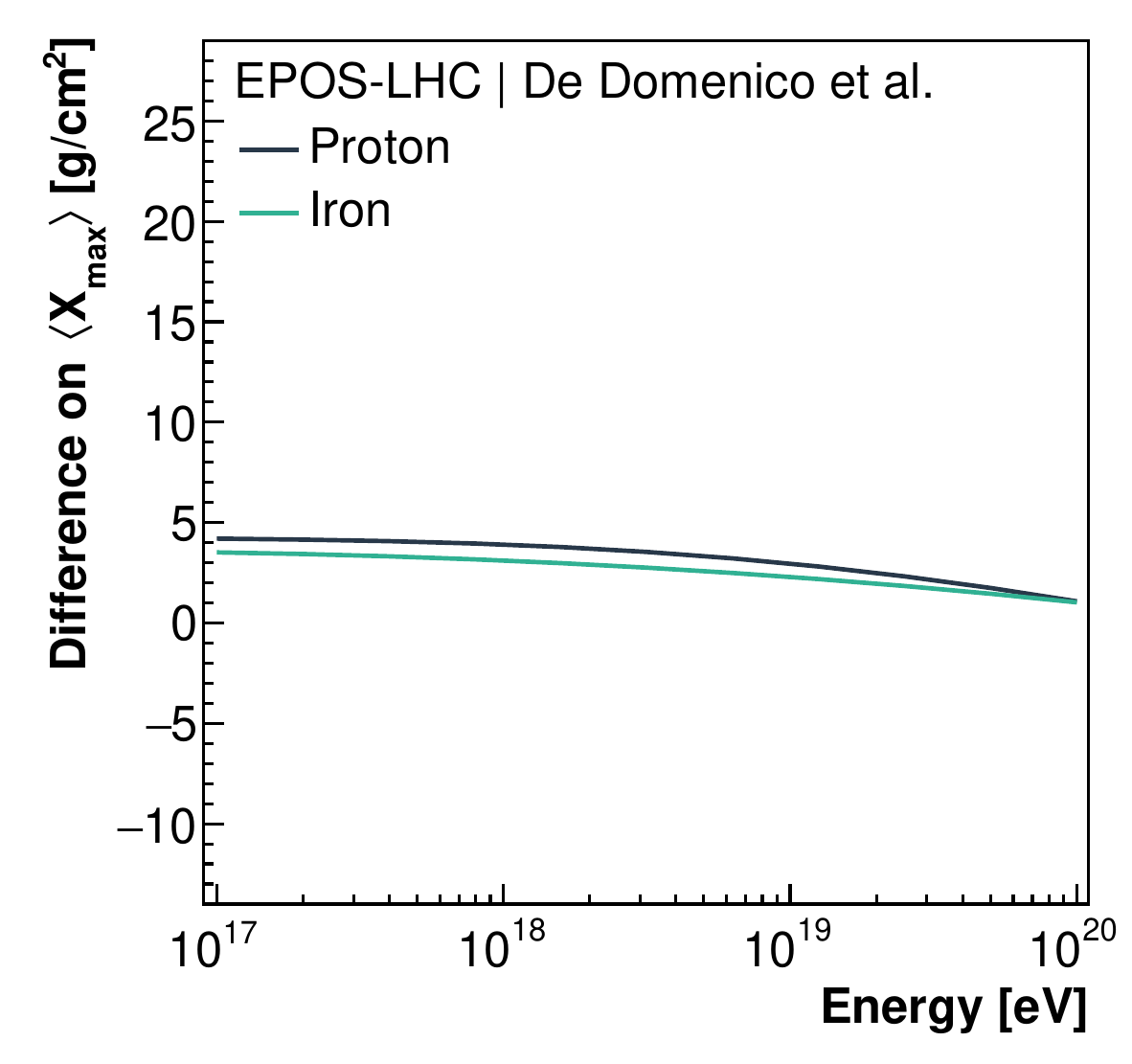} \\
  \includegraphics[width=0.32\textwidth]{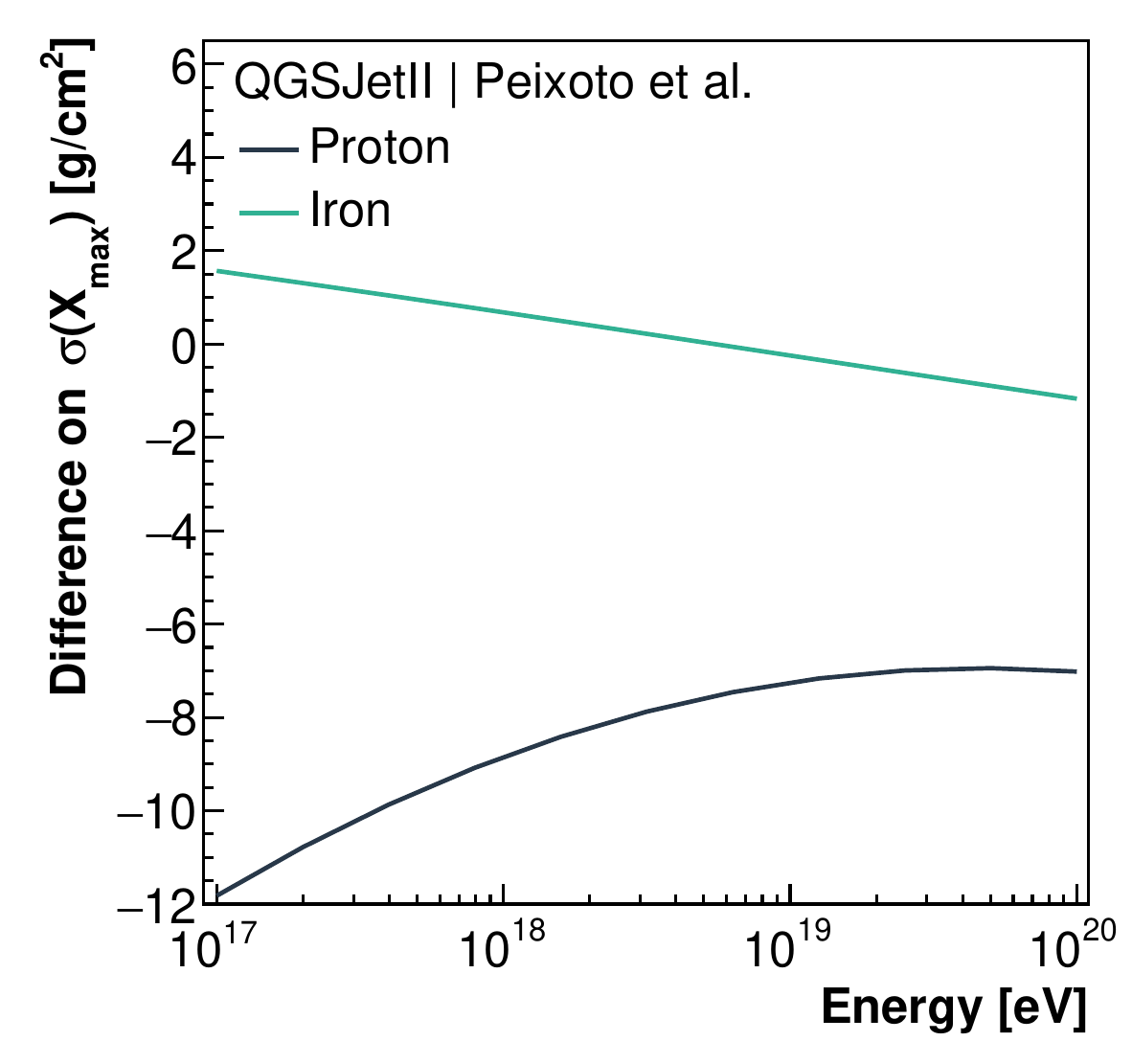}
  \includegraphics[width=0.32\textwidth]{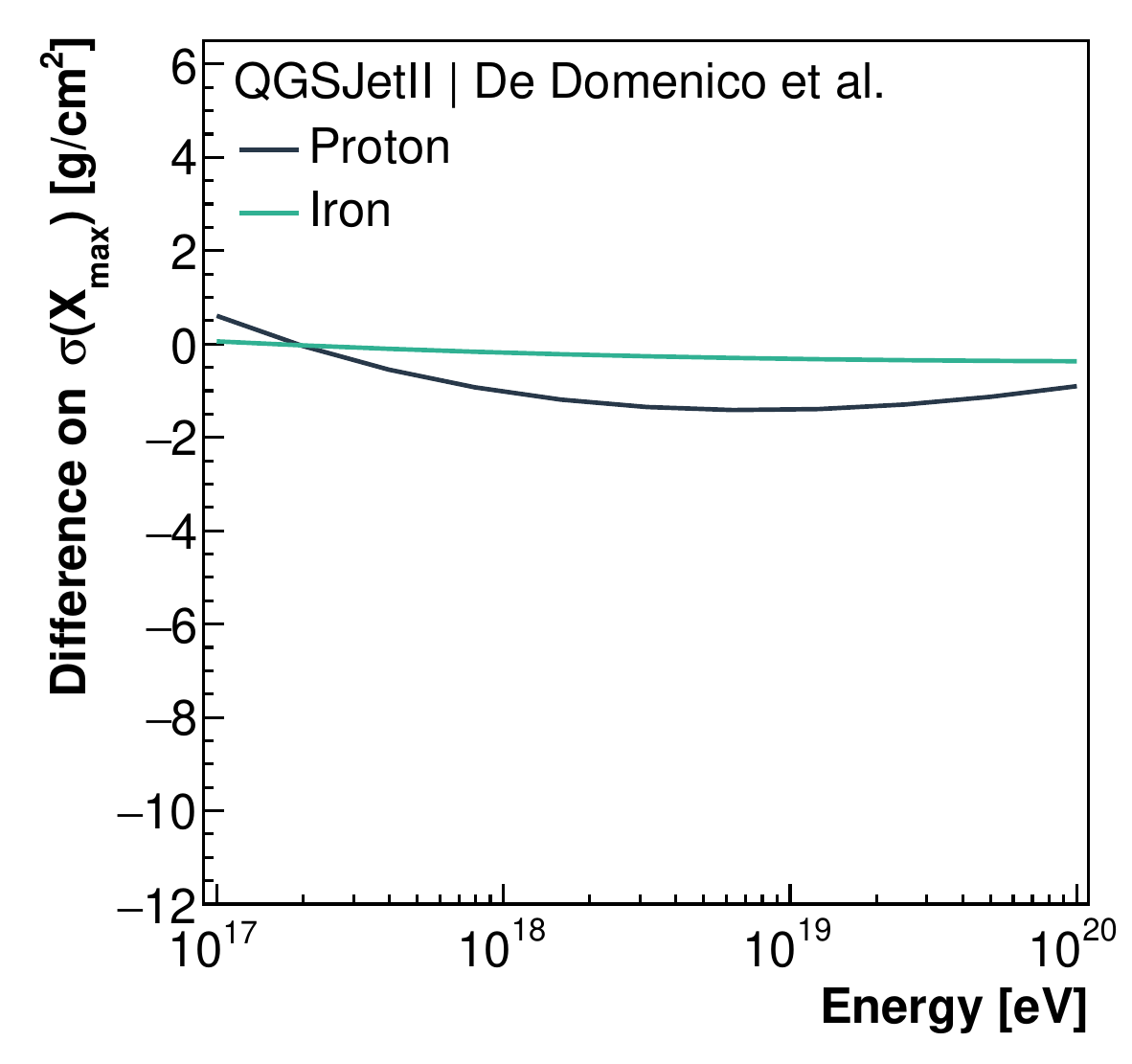}
  \includegraphics[width=0.32\textwidth]{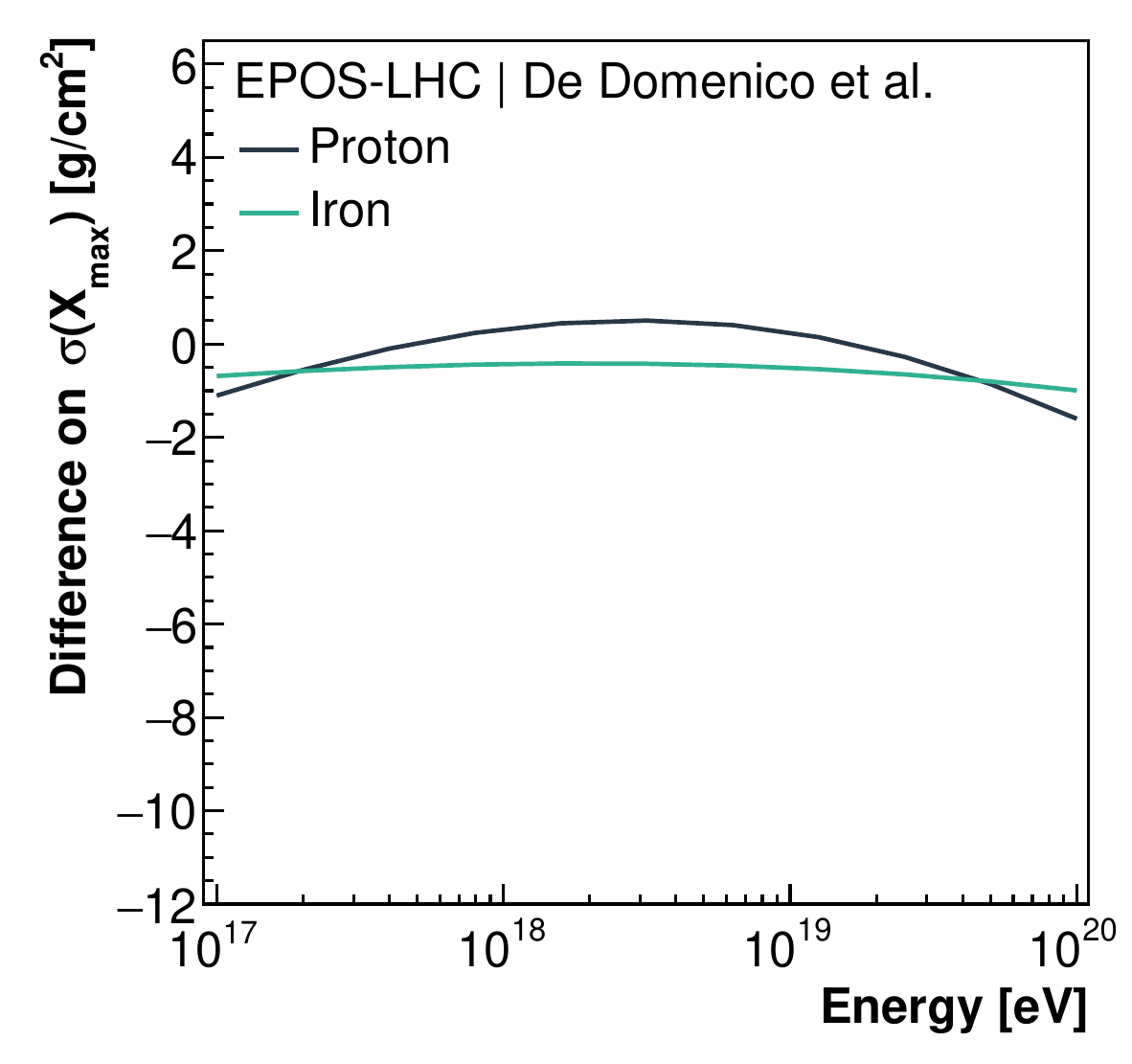}
  \caption{Comparison between first (upper plots) and second (lower plots) moments of parametrized \xmax distributions with parametrizations from \cite{xmax1} (left) and \cite{xmax2} (middle and right). Results are shown only for parametrizations based on simulations with the same hadronic model, which is indicated in the top-left corner of each plot.}
  \label{fig:compare-to-others}
\end{figure}

References \cite{xmax1} (Peixoto et al.) and \cite{xmax2} (de Domenico et al.) also proposed parametrizations of the \xmax distributions. The comparison of these parametrizations with the ones presented here is meaningful only when the same hadronic interaction models was used. Figure~\ref{fig:compare-to-others} compares first and second moments of the previously proposed parametrizations with the ones presented in this paper for simulations performed with \qgsjet and \epos hadronic interaction models. Previous parametrizations differs on the \meanXmax by as much as $20\,$g/cm$^2$ and on \sigmaXmax by as much as $12\,$g/cm$^2$.

\section{Conclusion}
\label{sec:conclusion}

The \xmax distribution is of great importance in UHECR studies and some functional forms have been proposed to describe it. In this paper, for the first time, three functions have been selected, explained, and compared to simulated \xmax distributions. A large sample of showers (10$^6$) has been generated for each point in a wide range of parameters space: four atomic nuclei have been considered: proton, carbon, silicon and iron with energies ranging from $10^{17}\,$eV to $10^{20}\,$eV and three hadronic interaction models: \epos, \qgsjet, and \sibyll. The primaries have also been mixed with fractions given by the best description of the Pierre Auger Observatory data.

In total three functions have been tested. Two were taken from the literature: Generalized Gumbel distribution~\cite{xmax2} and Exponentially Modified Gaussian distribution~\cite{xmax1}, and, in addition, the Log-normal distribution has been used as well. All functions have three parameters. The parameters have been fitted to the simulated \xmax distributions and the result is shown in table~\ref{tab:params}. The excellent quality of the fits allows the prediction of the first and second moments of the \xmax distribution with a maximum error of 2 and $3\,$g/cm$^2$, respectively.

The function that shows an overall best description of the \xmax distributions is the Generalized Gumbel distribution, followed by the Log-normal distribution. In some specific cases, the Log-normal distribution has a slightly better fit to the simulated distributions. However, in many other cases, the Generalized Gumbel distribution is much better than the Log-normal distribution. In studies of measured \xmax distribution, it is not possible to know beforehand which is the primary particle. Moreover, the hadronic interaction model dependence of the analysis must be minimized. For those reasons, the Generalized Gumbel distribution is proposed here as the best choice because it shows the best description for most of the cases. The Exponentially Modified Gaussian distribution is the one which most poorly describes the simulated showers among the three functions studied for almost all cases.

\section*{Acknowledgments}

Authors acknowledge FAPESP Project 2015/15897-1. VdS acknowledge CNPq. This study was financed in part by the Coordenação de Aperfeiçoamento de Pessoal de Nível Superior - Brasil (CAPES) - Finance Code 001. Authors acknowledge the National Laboratory for Scientific Computing (LNCC/MCTI, Brazil) for providing HPC resources of the SDumont supercomputer (http://sdumont.lncc.br). Thanks to Roger Clay and Michael Unger for reading and commenting the manuscript.

%
%
\begin{appendices}

%
%
\section{Akaike Information Criterion}
\label{app:aic}

The Akaike Information Criterion (AIC)~\cite{aic,burnham} is defined as

\begin{equation}
    \text{AIC} = 2k - 2\log \left( \mathcal{L}(\vec \theta) \right) \, ,
\end{equation}

\noindent where $k$ is the number of fitted parameters and $\mathcal{L}$ is the maximized value of the likelihood function for the fitted parameter set $\vec \theta$. Given a set of models, the AIC criterion calculates the relative quality of each model in describing the data. From its definition, it is expected that the model with the smaller AIC value in a set is the one that has the smallest statistical distance to the data set used to fit model parameters. 

Given that the absolute values of AIC carry no meaning and depend on the sample size, in this paper, model selection is based on the computation of the so-called Akaike differences ($\Delta_i$), or relative AIC. These are defined as the AIC values of each model with respect to the model with the smaller AIC, that is,

\begin{equation}
    \Delta_i = \text{AIC}_i - \text{AIC}_\text{min} \, .
\end{equation}

The relative AIC values provide a strong support for model selection in terms of information theory. A value of $\Delta_i = 0$ means that a model is preferred among the set to describe fitted data. Small values of $\Delta_i$ indicate that this model is not the best for the particular data set, but is competitive and should not be discarded. Large values of $\Delta_i$, on the other hand provide a strong argument against the $i$-th model.

%
%
\section{Parameter values}
\label{app:param}

In section~\ref{sec:parametrization} a method to describe the evolution of $\xmax$ distributions energy and mass is proposed in terms of equations~\ref{eq:p1} and~\ref{eq:p2}. This appendix compiles fitted parameters $a_i$, $b_i$ and $c_i$ of equation~\ref{eq:p2}. Table~\ref{tab:params} shows the values of fitted parameters whereas table~\ref{tab:errors}.

\begin{table}
  \centering
  \caption{Fitted parameters of equation~\ref{eq:p2} describing the evolution of $\xmax$ distributions as a function of primary energy and mass. Details in section~\ref{sec:parametrization}.}
  \scriptsize
  \begin{tabular}{c R{1.4cm} R{1.4cm} R{1.4cm} R{1.4cm} R{1.4cm} R{1.4cm} R{1.4cm} R{1.4cm} R{1.4cm}}
  \hline
\hline
\multicolumn{10}{c}{Exponentially modified gaussian} \\
\hline
   QGSJetII.04 &    $a_1$ &    $a_2$ &    $a_3$ &    $b_1$ &    $b_2$ &    $b_3$ &    $c_1$ &    $c_2$ &    $c_3$ \\
\hline
     $\lambda$ &    391.6 &  -354.39 &    97.21 &  -31.848 &   31.654 &   -9.193 &   0.7526 &  -0.7955 &   0.2407 \\
         $\mu$ &   -544.3 &  -152.02 &   -33.81 &   76.067 &   17.644 &    1.251 &  -0.4692 &  -0.5436 &        0 \\
      $\sigma$ &     44.9 &     -2.7 &    -4.35 &    -1.03 &     0.25 &        0 &        0 &        0 &        0 \\

\hline
      EPOS-LHC &    $a_1$ &    $a_2$ &    $a_3$ &    $b_1$ &    $b_2$ &    $b_3$ &    $c_1$ &    $c_2$ &    $c_3$ \\
\hline
     $\lambda$ &   478.15 &  -576.14 &   204.98 &  -41.845 &   55.499 &  -20.799 &   1.0336 &  -1.4561 &   0.5626 \\
         $\mu$ &  -757.99 &  -133.86 &   -32.58 &   99.306 &   15.373 &    0.914 &   -1.053 &  -0.4468 &        0 \\
      $\sigma$ &   239.07 &   -50.64 &   -18.25 &   -23.27 &    7.411 &    0.778 &    0.624 &  -0.2493 &        0 \\

\hline
    Sibyll2.3c &    $a_1$ &    $a_2$ &    $a_3$ &    $b_1$ &    $b_2$ &    $b_3$ &    $c_1$ &    $c_2$ &    $c_3$ \\
\hline
     $\lambda$ &    389.8 &  -199.83 &     3.46 &  -31.021 &    16.28 &    0.168 &   0.7148 &  -0.3993 &        0 \\
         $\mu$ &  -784.86 &    -3.33 &   -15.91 &  100.993 &   -0.983 &    0.433 &  -1.0377 &        0 &        0 \\
      $\sigma$ &    80.76 &     8.01 &    -10.2 &   -4.684 &   -0.815 &    0.526 &   0.0988 &        0 &        0 \\

\hline
\hline
\multicolumn{10}{c}{Generalized Gumbel} \\
\hline
   QGSJetII.04 &    $a_1$ &    $a_2$ &    $a_3$ &    $b_1$ &    $b_2$ &    $b_3$ &    $c_1$ &    $c_2$ &    $c_3$ \\
\hline
     $\lambda$ &     1.24 &    11.74 &    -6.85 &   -0.088 &   -1.393 &    0.855 &  0.00302 &  0.04702 & -0.02778 \\
         $\mu$ &  -368.79 &  -238.75 &   -32.14 &   61.443 &   25.159 &    1.255 &  -0.1138 &  -0.7326 &        0 \\
      $\sigma$ &     55.9 &     20.9 &    -15.9 &    -1.08 &     0.32 &        0 &        0 &        0 &        0 \\

\hline
      EPOS-LHC &    $a_1$ &    $a_2$ &    $a_3$ &    $b_1$ &    $b_2$ &    $b_3$ &    $c_1$ &    $c_2$ &    $c_3$ \\
\hline
     $\lambda$ &     4.34 &    -4.84 &     4.83 &  -0.4489 &    0.427 &   -0.314 &  0.01325 &        0 &        0 \\
         $\mu$ &  -565.11 &  -211.43 &   -36.32 &   82.199 &   22.453 &    1.288 &  -0.6189 &  -0.6475 &        0 \\
      $\sigma$ &    377.3 &      324 &   -228.1 &   -37.67 &   -29.63 &   22.436 &   1.0216 &   0.7366 &  -0.5955 \\

\hline
    Sibyll2.3c &    $a_1$ &    $a_2$ &    $a_3$ &    $b_1$ &    $b_2$ &    $b_3$ &    $c_1$ &    $c_2$ &    $c_3$ \\
\hline
     $\lambda$ &     0.02 &     0.16 &     0.11 &    0.038 &    0.014 &        0 &        0 &        0 &        0 \\
         $\mu$ &  -537.61 &  -131.99 &   -19.68 &   78.952 &   11.515 &    0.731 &  -0.4886 &  -0.3366 &        0 \\
      $\sigma$ &    60 &    24 &   -17 &    -1.06 &     -1.5 &     0.78 &        0 &        0 &        0 \\

\hline
\hline
\multicolumn{10}{c}{Log-normal} \\
\hline
   QGSJetII.04 &    $a_1$ &    $a_2$ &    $a_3$ &    $b_1$ &    $b_2$ &    $b_3$ &    $c_1$ &    $c_2$ &    $c_3$ \\
\hline
         $\mu$ &    8.974 &    -0.84 &    0.317 &  -0.3978 &   0.0684 &  -0.0344 &   0.0096 &        0 &        0 \\
      $\sigma$ &    0.532 &    -0.08 &    -0.04 &  -0.0065 &    -0.01 &   0.0066 &        0 &        0 &        0 \\
           $m$ &  -1152.4 &     64.9 &      -50 &   129.78 &    -8.74 &     4.65 &   -1.846 &        0 &        0 \\

\hline
      EPOS-LHC &    $a_1$ &    $a_2$ &    $a_3$ &    $b_1$ &    $b_2$ &    $b_3$ &    $c_1$ &    $c_2$ &    $c_3$ \\
\hline
         $\mu$ &   14.745 &    -2.04 &    0.043 &   -1.058 &   0.2761 &   -0.023 &  0.02806 & -0.00752 &        0 \\
      $\sigma$ &    0.034 &    -0.31 &   0.0043 &   0.0544 & -0.00104 &  0.00582 &-0.001768 &        0 &        0 \\
           $m$ &  -1745.1 &   -168.6 &     23.9 &   198.41 &     4.04 &     0.64 &   -3.738 &        0 &        0 \\

\hline
    Sibyll2.3c &    $a_1$ &    $a_2$ &    $a_3$ &    $b_1$ &    $b_2$ &    $b_3$ &    $c_1$ &    $c_2$ &    $c_3$ \\
\hline
         $\mu$ &     5.78 &    -0.03 &   -0.059 &  -0.0422 &  -0.0094 &        0 &        0 &        0 &        0 \\
      $\sigma$ &    0.551 &   -0.151 &    0.014 &  -0.0086 &   0.0026 &        0 &        0 &        0 &        0 \\
           $m$ &  -1085.3 &    -66.4 &     24.2 &   120.33 &     2.68 &    -1.38 &   -1.493 &        0 &        0 \\

\hline
\hline
  \end{tabular}
  \label{tab:params}
\end{table}

\begin{table}
  \centering
  \caption{Statistical uncertainty on fitted parameters of equation~\ref{eq:p2} describing the evolution of $\xmax$ distributions as a function of primary energy and mass. Details in section~\ref{sec:parametrization}.}
  \scriptsize
  \begin{tabular}{c R{1.4cm} R{1.4cm} R{1.4cm} R{1.4cm} R{1.4cm} R{1.4cm} R{1.4cm} R{1.4cm} R{1.4cm}}
    \hline
\hline
\multicolumn{10}{c}{Exponentially modified gaussian} \\
\hline
   QGSJetII.04 &    $a_1$ &    $a_2$ &    $a_3$ &    $b_1$ &    $b_2$ &    $b_3$ &    $c_1$ &    $c_2$ &    $c_3$ \\
\hline
     $\lambda$ &      0.1 &     0.08 &     0.05 &    0.006 &    0.004 &    0.003 &   0.0003 &   0.0002 &   0.0001 \\
         $\mu$ &      0.1 &     0.08 &     0.05 &    0.006 &    0.005 &    0.003 &   0.0003 &   0.0002 &        0 \\
      $\sigma$ &      0.4 &      0.3 &     0.03 &     0.02 &     0.02 &        0 &        0 &        0 &        0 \\

\hline
      EPOS-LHC &    $a_1$ &    $a_2$ &    $a_3$ &    $b_1$ &    $b_2$ &    $b_3$ &    $c_1$ &    $c_2$ &    $c_3$ \\
\hline
     $\lambda$ &     0.08 &     0.06 &     0.04 &    0.005 &    0.003 &    0.002 &   0.0002 &   0.0002 &   0.0001 \\
         $\mu$ &     0.08 &     0.06 &     0.04 &    0.005 &    0.004 &    0.002 &   0.0002 &   0.0002 &        0 \\
      $\sigma$ &     0.05 &     0.04 &     0.02 &    0.003 &    0.002 &    0.001 &   0.0002 &   0.0001 &        0 \\

\hline
    Sibyll2.3c &    $a_1$ &    $a_2$ &    $a_3$ &    $b_1$ &    $b_2$ &    $b_3$ &    $c_1$ &    $c_2$ &    $c_3$ \\
\hline
     $\lambda$ &      0.1 &     0.08 &     0.04 &    0.006 &    0.004 &    0.002 &   0.0003 &   0.0002 &        0 \\
         $\mu$ &      0.1 &     0.07 &     0.04 &    0.006 &    0.004 &    0.002 &   0.0003 &        0 &        0 \\
      $\sigma$ &     0.07 &     0.05 &     0.03 &    0.004 &    0.003 &    0.002 &   0.0002 &        0 &        0 \\

\hline
\hline
\multicolumn{10}{c}{Generalized Gumbel} \\
\hline
   QGSJetII.04 &    $a_1$ &    $a_2$ &    $a_3$ &    $b_1$ &    $b_2$ &    $b_3$ &    $c_1$ &    $c_2$ &    $c_3$ \\
\hline
     $\lambda$ &     0.01 &     0.03 &     0.02 &   0.0008 &    0.002 &    0.001 &    0.00003 &    0.00009 &    0.00006 \\
         $\mu$ &     0.09 &     0.06 &     0.04 &    0.005 &    0.004 &    0.002 &   0.0003 &   0.0002 &        0 \\
      $\sigma$ &      0.9 &      0.8 &      0.1 &     0.05 &     0.04 &        0 &        0 &        0 &        0 \\

\hline
      EPOS-LHC &    $a_1$ &    $a_2$ &    $a_3$ &    $b_1$ &    $b_2$ &    $b_3$ &    $c_1$ &    $c_2$ &    $c_3$ \\
\hline
     $\lambda$ &     0.01 &     0.04 &     0.03 &   0.0009 &    0.002 &    0.001 &    0.00003 &        0 &        0 \\
         $\mu$ &     0.07 &     0.05 &     0.03 &    0.004 &    0.003 &    0.002 &   0.0002 &   0.0001 &        0 \\
      $\sigma$ &      0.2 &      0.2 &      0.1 &     0.02 &     0.01 &    0.008 &   0.0007 &   0.0006 &   0.0004 \\

\hline
    Sibyll2.3c &    $a_1$ &    $a_2$ &    $a_3$ &    $b_1$ &    $b_2$ &    $b_3$ &    $c_1$ &    $c_2$ &    $c_3$ \\
\hline
     $\lambda$ &     0.03 &     0.04 &    0.007 &    0.001 &    0.002 &        0 &        0 &        0 &        0 \\
         $\mu$ &     0.08 &     0.06 &     0.04 &    0.005 &    0.004 &    0.002 &   0.0002 &   0.0002 &        0 \\
      $\sigma$ &        1 &        3 &        2 &     0.06 &      0.1 &     0.08 &        0 &        0 &        0 \\

\hline
\hline
\multicolumn{10}{c}{Log-normal} \\
\hline
   QGSJetII.04 &    $a_1$ &    $a_2$ &    $a_3$ &    $b_1$ &    $b_2$ &    $b_3$ &    $c_1$ &    $c_2$ &    $c_3$ \\
\hline
         $\mu$ &    0.006 &    0.007 &    0.004 &   0.0004 &   0.0004 &   0.0002 &    0.00002 &        0 &        0 \\
      $\sigma$ &    0.007 &     0.02 &     0.01 &   0.0004 &    0.001 &   0.0006 &        0 &        0 &        0 \\
           $m$ &      0.7 &      0.8 &      0.5 &     0.05 &     0.04 &     0.03 &    0.002 &        0 &        0 \\

\hline
      EPOS-LHC &    $a_1$ &    $a_2$ &    $a_3$ &    $b_1$ &    $b_2$ &    $b_3$ &    $c_1$ &    $c_2$ &    $c_3$ \\
\hline
         $\mu$ &    0.006 &    0.007 &    0.004 &   0.0004 &   0.0004 &   0.0002 &    0.00002 &   0.00002 &        0 \\
      $\sigma$ &    0.002 &    0.001 &   0.0009 &   0.0001 &  0.00007 &    0.00005 &    0.000005 &        0 &        0 \\
           $m$ &      0.6 &      0.7 &      0.4 &     0.04 &     0.04 &     0.02 &    0.002 &        0 &        0 \\

\hline
    Sibyll2.3c &    $a_1$ &    $a_2$ &    $a_3$ &    $b_1$ &    $b_2$ &    $b_3$ &    $c_1$ &    $c_2$ &    $c_3$ \\
\hline
         $\mu$ &     0.02 &     0.02 &    0.002 &   0.0009 &   0.0008 &        0 &        0 &        0 &        0 \\
      $\sigma$ &    0.007 &    0.005 &   0.0007 &   0.0004 &   0.0003 &        0 &        0 &        0 &        0 \\
           $m$ &      0.6 &      0.5 &      0.3 &     0.04 &     0.03 &     0.02 &    0.002 &        0 &        0 \\

\hline
\hline

  \end{tabular}
  \label{tab:errors}
\end{table}

%
%
\section{Updated parametrization}
\label{app:update}

This section presents a parametrization analogous to that of appendix~\ref{app:param}, with the difference that helium-initiated showers were included in the simulation library used in the fit procedure. Table~\ref{tab:params:update} presents the fitted parameters and table~\ref{tab:errors:update} shows the statistical uncertainties associated the parameters.

\begin{table}
    \centering
    \caption{Coefficients of the updated parametrization.}
    \scriptsize
    \begin{tabular}{c R{1.4cm} R{1.4cm} R{1.4cm} R{1.4cm} R{1.4cm} R{1.4cm} R{1.4cm} R{1.4cm} R{1.4cm}}
    \hline
    \hline
    \multicolumn{10}{c}{Exponentially modified gaussian} \\
    \hline
     QGSJetII.04 &        $a_1$ &        $a_2$ &        $a_3$ &        $b_1$ &        $b_2$ &        $b_3$ &        $c_1$ &        $c_2$ &        $c_3$ \\
    \hline
       $\lambda$ & $    383.46$ & $   -356.01$ & $    101.44$ & $   -30.988$ & $    31.724$ & $    -9.579$ & $    0.7286$ & $  -0.79657$ & $   0.25094$ \\
           $\mu$ & $   -574.89$ & $     -8.43$ & $   -111.97$ & $    79.455$ & $     2.183$ & $     9.633$ & $   -0.5610$ & $  -0.12668$ & $  -0.22589$ \\
        $\sigma$ & $    156.67$ & $    -60.67$ & $      0.29$ & $   -13.056$ & $     6.568$ & $   -0.5517$ & $    0.3234$ & $  -0.16979$ & $   0.01489$ \\
    \hline
        EPOS-LHC &        $a_1$ &        $a_2$ &        $a_3$ &        $b_1$ &        $b_2$ &        $b_3$ &        $c_1$ &        $c_2$ &        $c_3$ \\
    \hline
       $\lambda$ & $    467.66$ & $   -570.44$ & $    205.90$ & $   -40.486$ & $    54.771$ & $  -20.9147$ & $    0.9920$ & $  -1.43480$ & $   0.56646$ \\
           $\mu$ & $   -751.27$ & $    -35.75$ & $    -96.00$ & $    98.577$ & $     4.680$ & $    7.8246$ & $  -1.03258$ & $  -0.15658$ & $  -0.18779$ \\
        $\sigma$ & $    261.54$ & $    -98.75$ & $      3.18$ & $   -25.563$ & $    12.681$ & $   -1.6304$ & $   0.68366$ & $  -0.39291$ & $   0.06668$ \\
    \hline
      Sibyll2.3c &        $a_1$ &        $a_2$ &        $a_3$ &        $b_1$ &        $b_2$ &        $b_3$ &        $c_1$ &        $c_2$ &        $c_3$ \\
    \hline
       $\lambda$ & $    400.21$ & $   -298.36$ & $     61.12$ & $   -32.122$ & $    26.888$ & $    -6.048$ & $    0.7421$ & $  -0.68589$ & $   0.16888$ \\
           $\mu$ & $   -762.86$ & $    133.05$ & $   -115.19$ & $    98.588$ & $   -15.634$ & $   11.1213$ & $   -0.9698$ & $   0.39507$ & $  -0.28936$ \\
        $\sigma$ & $    103.77$ & $     17.05$ & $    -26.37$ & $    -7.208$ & $    -1.774$ & $    2.2770$ & $   0.16829$ & $   0.02605$ & $  -0.04786$ \\
    \hline
    \hline
    \multicolumn{10}{c}{Generalized Gumbel} \\
    \hline
     QGSJetII.04 &        $a_1$ &        $a_2$ &        $a_3$ &        $b_1$ &        $b_2$ &        $b_3$ &        $c_1$ &        $c_2$ &        $c_3$ \\
    \hline
       $\lambda$ & $    0.6465$ & $     1.932$ & $     1.767$ & $  -0.02109$ & $   -0.2143$ & $  -0.15658$ & $  0.001274$ & $  0.008431$ & $  0.004271$ \\
           $\mu$ & $   -412.77$ & $   -136.12$ & $   -76.789$ & $   66.4057$ & $   13.8397$ & $    6.1521$ & $  -0.24926$ & $  -0.42961$ & $  -0.12990$ \\
        $\sigma$ & $    211.50$ & $    -13.05$ & $    -13.36$ & $   -17.650$ & $    2.1148$ & $    0.8359$ & $   0.44432$ & $  -0.03287$ & $  -0.02996$ \\
    \hline
        EPOS-LHC &        $a_1$ &        $a_2$ &        $a_3$ &        $b_1$ &        $b_2$ &        $b_3$ &        $c_1$ &        $c_2$ &        $c_3$ \\
    \hline
       $\lambda$ & $     5.621$ & $     4.614$ & $    -2.730$ & $   -0.5942$ & $   -0.5019$ & $    0.4506$ & $  0.017490$ & $   0.01506$ & $  -0.01395$ \\
           $\mu$ & $   -505.56$ & $   -356.03$ & $     27.79$ & $    75.805$ & $    38.127$ & $   -5.6766$ & $  -0.44325$ & $  -1.08321$ & $   0.19476$ \\
        $\sigma$ & $    446.23$ & $   -209.52$ & $     61.24$ & $   -45.209$ & $    27.790$ & $    -8.548$ & $    1.2330$ & $  -0.86067$ & $   0.26753$ \\
    \hline
      Sibyll2.3c &        $a_1$ &        $a_2$ &        $a_3$ &        $b_1$ &        $b_2$ &        $b_3$ &        $c_1$ &        $c_2$ &        $c_3$ \\
    \hline
       $\lambda$ & $    -1.164$ & $     5.295$ & $    -1.288$ & $    0.1623$ & $   -0.5541$ & $    0.1595$ & $ -0.003128$ & $  0.015849$ & $ -0.004691$ \\
           $\mu$ & $   -608.46$ & $     58.10$ & $   -111.29$ & $    86.634$ & $    -9.002$ & $    10.606$ & $   -0.6925$ & $   0.21420$ & $  -0.26591$ \\
        $\sigma$ & $    123.53$ & $    100.69$ & $    -64.16$ & $    -8.066$ & $    -9.734$ & $    5.9116$ & $    0.1955$ & $   0.21773$ & $  -0.13790$ \\
    \hline
    \hline
    \multicolumn{10}{c}{Log-normal} \\
    \hline
     QGSJetII.04 &        $a_1$ &        $a_2$ &        $a_3$ &        $b_1$ &        $b_2$ &        $b_3$ &        $c_1$ &        $c_2$ &        $c_3$ \\
    \hline
           $\mu$ & $    7.4815$ & $    4.9797$ & $   -3.7692$ & $  -0.23720$ & $  -0.53973$ & $  0.394916$ & $ 0.0053502$ & $ 0.0150842$ & $-0.0107345$ \\
        $\sigma$ & $    1.2503$ & $   -2.3907$ & $    1.1886$ & $  -0.08393$ & $   0.23658$ & $ -0.124054$ & $ 0.0020535$ & $-0.0063788$ & $ 0.0033612$ \\
             $m$ & $   -838.51$ & $   -882.00$ & $    458.23$ & $    95.812$ & $    91.716$ & $   -49.171$ & $   -0.9373$ & $  -2.56672$ & $   1.36078$ \\
    \hline
        EPOS-LHC &        $a_1$ &        $a_2$ &        $a_3$ &        $b_1$ &        $b_2$ &        $b_3$ &        $c_1$ &        $c_2$ &        $c_3$ \\
    \hline
           $\mu$ & $   14.6051$ & $   -9.7557$ & $    2.9011$ & $  -1.04208$ & $   1.13565$ & $ -0.350791$ & $  0.027637$ & $-0.0321144$ & $ 0.0098292$ \\
        $\sigma$ & $   -0.1712$ & $    0.9208$ & $   -0.1666$ & $   0.07662$ & $  -0.13820$ & $  0.027235$ & $-0.0023838$ & $ 0.0039718$ & $-0.0007602$ \\
             $m$ & $  -1712.52$ & $    511.54$ & $   -184.49$ & $   194.863$ & $   -70.418$ & $   23.9702$ & $  -3.64482$ & $   2.11658$ & $  -0.70430$ \\
    \hline
      Sibyll2.3c &        $a_1$ &        $a_2$ &        $a_3$ &        $b_1$ &        $b_2$ &        $b_3$ &        $c_1$ &        $c_2$ &        $c_3$ \\
    \hline
           $\mu$ & $    7.2028$ & $    0.5040$ & $   -1.0804$ & $  -0.19828$ & $  -0.06398$ & $   0.10976$ & $  0.004289$ & $  0.001508$ &  $-0.003020$ \\
        $\sigma$ & $    1.0921$ & $   -0.7835$ & $    0.3047$ & $  -0.06781$ & $   0.07805$ & $ -0.035924$ & $ 0.0015989$ & $-0.0022613$ & $ 0.0011171$ \\
             $m$ & $   -985.13$ & $    -27.55$ & $     11.50$ & $   109.434$ & $     0.521$ & $    -1.363$ & $   -1.2006$ & $  -0.00453$ & $   0.04163$ \\
    \hline
    \hline
    \end{tabular}
    \label{tab:params:update}
\end{table}

\begin{table}
    \centering
    \caption{Statistical uncertainties associated to the new set of coefficients.}
    \scriptsize
    \begin{tabular}{c R{1.4cm} R{1.4cm} R{1.4cm} R{1.4cm} R{1.4cm} R{1.4cm} R{1.4cm} R{1.4cm} R{1.4cm}}
    \hline
    \hline
    \multicolumn{10}{c}{Exponentially modified gaussian} \\
    \hline
     QGSJetII.04 &        $a_1$ &        $a_2$ &        $a_3$ &        $b_1$ &        $b_2$ &        $b_3$ &        $c_1$ &        $c_2$ &        $c_3$ \\
    \hline
       $\lambda$ & $      0.04$ & $      0.03$ & $      0.02$ & $     0.002$ & $     0.002$ & $     0.001$ & $    0.0001$ & $   0.00009$ & $   0.00005$ \\
           $\mu$ & $      0.04$ & $      0.03$ & $      0.02$ & $     0.002$ & $     0.002$ & $     0.001$ & $    0.0001$ & $   0.00009$ & $   0.00005$ \\
        $\sigma$ & $      0.04$ & $      0.03$ & $      0.02$ & $     0.002$ & $     0.002$ & $    0.0009$ & $    0.0001$ & $   0.00008$ & $   0.00005$ \\
    \hline
        EPOS-LHC &        $a_1$ &        $a_2$ &        $a_3$ &        $b_1$ &        $b_2$ &        $b_3$ &        $c_1$ &        $c_2$ &        $c_3$ \\
    \hline
       $\lambda$ & $      0.03$ & $      0.02$ & $      0.02$ & $     0.002$ & $     0.001$ & $    0.0009$ & $    0.0001$ & $   0.00007$ & $   0.00004$ \\
           $\mu$ & $      0.03$ & $      0.02$ & $      0.02$ & $     0.002$ & $     0.001$ & $    0.0008$ & $   0.00009$ & $   0.00007$ & $   0.00004$ \\
        $\sigma$ & $      0.03$ & $      0.02$ & $      0.01$ & $     0.002$ & $     0.001$ & $    0.0008$ & $   0.00008$ & $   0.00006$ & $   0.00004$ \\
    \hline
      Sibyll2.3c &        $a_1$ &        $a_2$ &        $a_3$ &        $b_1$ &        $b_2$ &        $b_3$ &        $c_1$ &        $c_2$ &        $c_3$ \\
    \hline
       $\lambda$ & $      0.04$ & $      0.03$ & $      0.02$ & $     0.002$ & $     0.002$ & $     0.001$ & $    0.0001$ & $   0.00008$ & $   0.00005$ \\
           $\mu$ & $      0.03$ & $      0.03$ & $      0.02$ & $     0.002$ & $     0.001$ & $    0.0009$ & $    0.0001$ & $   0.00007$ & $   0.00005$ \\
        $\sigma$ & $      0.03$ & $      0.02$ & $      0.01$ & $     0.002$ & $     0.001$ & $    0.0009$ & $   0.00008$ & $   0.00007$ & $   0.00004$ \\
    \hline
    \hline
    \multicolumn{10}{c}{Generalized Gumbel} \\
    \hline
     QGSJetII.04 &        $a_1$ &        $a_2$ &        $a_3$ &        $b_1$ &        $b_2$ &        $b_3$ &        $c_1$ &        $c_2$ &        $c_3$ \\
    \hline
       $\lambda$ & $    0.0009$ & $     0.002$ & $     0.001$ & $   0.00005$ & $    0.0001$ & $   0.00007$ & $  0.000003$ & $  0.000006$ & $  0.000005$ \\
           $\mu$ & $      0.02$ & $      0.01$ & $     0.008$ & $    0.0009$ & $    0.0007$ & $    0.0004$ & $   0.00005$ & $   0.00004$ & $   0.00002$ \\
        $\sigma$ & $      0.02$ & $      0.03$ & $      0.01$ & $     0.001$ & $    0.0008$ & $    0.0006$ & $   0.00005$ & $   0.00005$ & $   0.00006$ \\
    \hline
        EPOS-LHC &        $a_1$ &        $a_2$ &        $a_3$ &        $b_1$ &        $b_2$ &        $b_3$ &        $c_1$ &        $c_2$ &        $c_3$ \\
    \hline
       $\lambda$ & $     0.002$ & $     0.004$ & $     0.004$ & $    0.0001$ & $    0.0003$ & $    0.0002$ & $  0.000006$ & $   0.00001$ & $   0.00001$ \\
           $\mu$ & $      0.03$ & $      0.02$ & $      0.01$ & $     0.002$ & $     0.001$ & $    0.0009$ & $   0.00009$ & $   0.00007$ & $   0.00004$ \\
        $\sigma$ & $      0.04$ & $      0.03$ & $      0.02$ & $     0.002$ & $     0.002$ & $     0.001$ & $    0.0001$ & $   0.00009$ & $   0.00006$ \\
    \hline
      Sibyll2.3c &        $a_1$ &        $a_2$ &        $a_3$ &        $b_1$ &        $b_2$ &        $b_3$ &        $c_1$ &        $c_2$ &        $c_3$ \\
    \hline
       $\lambda$ & $     0.002$ & $     0.002$ & $     0.002$ & $    0.0001$ & $    0.0001$ & $    0.0001$ & $  0.000006$ & $  0.000007$ & $  0.000005$ \\
           $\mu$ & $      0.04$ & $      0.03$ & $      0.02$ & $     0.002$ & $     0.002$ & $     0.001$ & $    0.0001$ & $   0.00008$ & $   0.00005$ \\
        $\sigma$ & $      0.03$ & $      0.03$ & $      0.02$ & $     0.002$ & $     0.001$ & $    0.0009$ & $    0.0001$ & $   0.00008$ & $   0.00005$ \\
    \hline
    \hline
    \multicolumn{10}{c}{Log-normal} \\
    \hline
     QGSJetII.04 &        $a_1$ &        $a_2$ &        $a_3$ &        $b_1$ &        $b_2$ &        $b_3$ &        $c_1$ &        $c_2$ &        $c_3$ \\
    \hline
           $\mu$ & $    0.0003$ & $    0.0003$ & $    0.0002$ & $   0.00002$ & $   0.00001$ & $  0.000009$ & $ 0.0000009$ & $ 0.0000007$ & $ 0.0000005$ \\
        $\sigma$ & $    0.0002$ & $    0.0002$ & $    0.0001$ & $   0.00001$ & $   0.00001$ & $  0.000006$ & $ 0.0000006$ & $ 0.0000005$ & $ 0.0000003$ \\
             $m$ & $      0.04$ & $      0.03$ & $      0.02$ & $     0.002$ & $     0.002$ & $     0.001$ & $    0.0001$ & $   0.00009$ & $   0.00005$ \\
    \hline
        EPOS-LHC &        $a_1$ &        $a_2$ &        $a_3$ &        $b_1$ &        $b_2$ &        $b_3$ &        $c_1$ &        $c_2$ &        $c_3$ \\
    \hline
           $\mu$ & $    0.0003$ & $    0.0002$ & $    0.0002$ & $   0.00002$ & $   0.00001$ & $  0.000009$ & $  0.000001$ & $ 0.0000007$ & $ 0.0000005$ \\
        $\sigma$ & $    0.0002$ & $    0.0002$ & $    0.0001$ & $   0.00001$ & $   0.00001$ & $  0.000006$ & $ 0.0000006$ & $ 0.0000005$ & $ 0.0000003$ \\
             $m$ & $      0.03$ & $      0.03$ & $      0.02$ & $     0.002$ & $     0.001$ & $    0.0008$ & $   0.00009$ & $   0.00007$ & $   0.00004$ \\
    \hline
      Sibyll2.3c &        $a_1$ &        $a_2$ &        $a_3$ &        $b_1$ &        $b_2$ &        $b_3$ &        $c_1$ &        $c_2$ &        $c_3$ \\
    \hline
           $\mu$ & $    0.0004$ & $    0.0003$ & $    0.0002$ & $   0.00002$ & $   0.00002$ & $   0.00001$ & $  0.000001$ & $  0.000001$ & $  0.000001$ \\
        $\sigma$ & $    0.0003$ & $    0.0002$ & $    0.0001$ & $   0.00002$ & $   0.00001$ & $  0.000008$ & $ 0.0000008$ & $ 0.0000006$ & $ 0.0000004$ \\
             $m$ & $      0.04$ & $      0.03$ & $      0.02$ & $     0.002$ & $     0.002$ & $     0.001$ & $    0.0001$ & $   0.00008$ & $   0.00005$ \\
    \hline
    \hline
    \end{tabular}
    \label{tab:errors:update}
\end{table}

%
%
\end{appendices}

\end{document}